\documentclass[12pt,dvips]{article}
\usepackage{rotating}
\usepackage{epsfig}
\usepackage{color}
\usepackage{cite}

\newcommand{\psl}{\mathbf{p} \hspace{-0.5 em}/}

\newcommand{\lsim}{\raisebox{-0.13cm}{~\shortstack{$<$ \\[-0.07cm] $\sim$}}~}
\newcommand{\gsim}{\raisebox{-0.13cm}{~\shortstack{$>$ \\[-0.07cm] $\sim$}}~}

\begin{document}

\def\thefootnote{\fnsymbol{footnote}}

\begin{flushright}
\end{flushright}

\begin{center}
{\bf {\LARGE
Measuring the Higgs boson mass with  \\[2mm]
transverse mass variables
} }
\end{center}

\medskip

\begin{center}{\large
Kiwoon~Choi$^a$,
Jae~Sik~Lee$^b$, and
Chan~Beom~Park$^a$
}
\end{center}

\begin{center}
{\em $^a$Department of Physics, KAIST, Daejeon 305-701, Korea}\\[0.2cm]
{\em $^b$Physics Division, National Center for Theoretical Sciences,
Hsinchu, Taiwan}\\[0.2cm]
\end{center}

\bigskip

\centerline{\bf ABSTRACT}
\begin{flushleft}
\medskip\noindent
We provide a comparative study of the Higgs boson mass measurements
based on two approaches to the dileptonic decay of $W$ bosons
produced by the Higgs boson decay, one using the kinematic variable
$M_T^{\rm true}$ and the other using the $M_{T2}$-assisted on-shell
reconstruction of the invisible neutrino momenta. We find that
these two approaches can determine the Higgs boson mass with a
similar accuracy for both of the two main production mechanisms of
the SM Higgs boson at the LHC, i.e. the gluon-gluon fusion and the
weak vector boson fusion. 
We also notice that the Higgs signal
distribution for the gluon-gluon fusion becomes narrower under the
$M_{T2}$ cut, while the corresponding  background distribution
becomes flatter, indicating that one might be able to reduce the
systematic uncertainties of mass measurement with an appropriate
$M_{T2}$ cut.

%
\end{flushleft}

\newpage

\section{Introduction}
\label{sec:intro}

The utmost target of the LHC is to discover the Higgs boson and study
its fundamental properties to establish the mechanism of electroweak
symmetry breaking as the origin of particle
masses~\cite{Aad:2009wy,Ball:2007zza}.
The unsuccessful search  at the LEP experiment set a lower bound on
the standard model (SM) Higgs mass at 114.4 GeV (95 \%
C.L.)~\cite{Barate:2003sz}, while the analysis of the electroweak
precision data indicates a relatively light SM Higgs boson with $m_H
\stackrel{<}{{}_\sim} 185$~GeV at the 95 \% confidence
level~\cite{Collaboration:2008ub}.
%
Combined with the recent Tevatron search excluding 158 GeV \,$\leq
m_H\leq$\, 175 GeV~\cite{TevatronHiggs}, we anticipate that the
SM Higgs boson most probably lies in the mass range between 114.4
GeV and 158 GeV.

At the LHC, the Higgs boson is mainly produced by the gluon-gluon
fusion (GGF) mechanism, and the second most important source is the
weak vector boson fusion (VBF)
process~\cite{Spira:1997dg,Djouadi:2005gi}. While the production
cross section of GGF is about 10 times larger than that of VBF for
the Higgs mass region $114.4 ~{\rm GeV} \leq m_H \leq 170~{\rm
GeV}$, VBF has its advantage in a kinematic structure containing two
forwarding tagging jets with a large rapidity gap, which can be
exploited to isolate the Higgs boson signal from backgrounds and to
study the signal properties~\cite{Asai:2004ws,Ruwiedel:2007zz,Yazgan:2007cd}.

The specific search strategy of the Higgs boson at the LHC depends
on its mass and decay pattern.  According to the combined study on
the expected Higgs discovery significance at ATLAS with 10 fb$^{-1}$
luminosity~\cite{Aad:2009wy}, when the Higgs boson is lighter than
 130 GeV,
the main search channel is GGF with the Higgs boson decaying into
two tau leptons or into two photons.
%
On the other hand, for $130~{\rm GeV} \lsim m_H \lsim 150$ GeV, the
following three channels are
available with a similar significance: $(i)$ GGF with the Higgs
boson decaying into two $Z$ bosons which subsequently decay into four
charged leptons, $(ii)$ GGF with the Higgs boson decaying into two
$W$ bosons, which subsequently decay into two charged leptons and two
neutrinos, and $(iii)$ VBF with the Higgs boson decaying into two
$W$ bosons, which subsequently decay into two charged leptons and two
neutrinos.
Finally, for $150~{\rm GeV} \lsim m_H \lsim 190$ GeV, Higgs boson
decay into two $Z$ bosons is relatively suppressed, while
channels $(ii)$ and $(iii)$ remain as the dominant
search channels~\cite{Barger:1990mn}.
Therefore, dileptonic $W$ boson decays play a crucial role in the
Higgs boson search at the LHC when the Higgs boson weighs between
130 GeV and 190 GeV. Furthermore, if one can improve the efficiency
of dileptonic channels, they might play an important role even for
$m_H$ lighter than 130 GeV.

In dileptonic $W$ boson decays, there are two invisible neutrinos
which make a direct reconstruction of the Higgs boson mass 
impossible. To overcome this difficulty, one can consider various kind
of transverse mass variables.\footnote{
The general
concept of transverse mass has been introduced and studied as early
as in Ref.~\cite{vanNeerven:1982mz}. In our work, we concentrate on
the recently proposed transverse mass variables such as
$M_T^{\rm true}$, $M_{T2}$, and $m_H^{\rm maos}$.
}
%
A well-known example
is the transverse mass of a $W$ boson pair in the process $H\rightarrow
WW\rightarrow \ell \nu \ell^\prime\nu^\prime$ ($\ell,\,\ell^\prime
=e,\,\mu$ and $\nu,\,\nu^\prime =\nu_e,\,\nu_\mu$):
\begin{equation}
M_T^2(WW)=m_{\ell\ell^\prime}^2+m_{\nu\nu^\prime}^2
+2\left(\sqrt{\left|{\bf
p}_T^{\ell\ell^\prime}\right|^2+m_{\ell\ell^\prime}^2}
\sqrt{\left|{\bf p}_T^{\nu\nu^\prime}\right|^2+m_{\nu\nu^\prime}^2}
-{\bf p}_T^{\ell\ell^\prime}\cdot {\bf
p}_T^{\nu\nu^\prime}\right)\,, \label{eq:mtww}
\end{equation}
where $m_X$ and ${\bf p}^X_T$  denote the invariant mass and the
transverse momentum, respectively, of $X=ll^\prime, \nu\nu^\prime$.
 Obviously $M_T(WW)$
is bounded by $m_H$, and this upper bound is saturated when
$ll^\prime$ and $\nu\nu^\prime$ have the same rapidity,
$\eta_{ll^\prime}=\eta_{\nu\nu^\prime}$, where
$\eta = \frac{1}{2}\ln (E+p_L)/(E-p_L)$ for the longitudinal
momentum $p_L$. 
Therefore, if $M_T(WW)$ is correctly
reconstructed event by event, the Higgs boson mass can be read off
from the endpoint value of $M_T(WW)$ distribution. However, there is
an obstacle to this approach: $m_{\nu\nu^\prime}$ cannot be
experimentally determined, although ${\bf p}_T^{\nu\nu^\prime}$ can
be deduced from the missing transverse momentum of the event.
One way to bypass this difficulty
%
is to simply take $m_{\nu\nu^\prime}=m_{ll^\prime}$
\cite{Rainwater:1999sd}, and consider the distribution of
\begin{eqnarray}
M_T^{\rm approx}\equiv
\left.M_T(WW)\right|_{m_{\nu\nu^\prime}=m_{\ell\ell^\prime}},\end{eqnarray}
which would provide a good approximation to the true value of
$M_T(WW)$ if the $W$ boson pair were produced at near-threshold.
However, in reality, a sizable number of events are not close to such
a threshold, and as a result,
$M_T^{\rm approx}$ is not
   bounded by $m_H$ anymore.
   Still, the detailed shape of its distribution has a certain correlation with $m_H$,
so $M_T^{\rm approx}$ has been widely used in the previous studies
of the Higgs boson search and mass
measurement~\cite{Aad:2009wy,Asai:2004ws,Yazgan:2007cd}.

   Recently, it has been noticed that an alternative transverse mass variable~\cite{Barr:2009mx},
   \begin{eqnarray}M_T^{\rm
   true}\equiv\left.M_T(WW)\right|_{m_{\nu\nu^\prime}=0},\end{eqnarray}
   may determine the Higgs boson mass more accurately than $M_T^{\rm approx}$
   does,
   since it is
   bounded by $m_H$ and thus could have a stronger correlation with $m_H$~\cite{Barr:2009mx,DPinNath:2010zj}.
Obviously, for each event
   \begin{equation}
   M_T^{\rm true}\,\leq\, M_T(WW) \,\leq\, m_H,
   \end{equation} and
  the upper bound of $M_T^{\rm true}$  is saturated by the events with
   $\eta_{ll^\prime}=\eta_{\nu\nu^\prime}$ and
   $m_{\nu\nu^\prime}=0$; therefore, the endpoint value of $M_T^{\rm
   true}$ distribution
   indeed corresponds to $m_H$.

There is another widely discussed transverse mass variable,
$M_{T2}$, which
is defined for a generic event
with two
identical invisible particles~\cite{refs:mt2} and was applied recently
to the mass measurement of supersymmetric particles~\cite{refs:mt2,mt2_kink}.
For the Higgs boson event $H\rightarrow WW\rightarrow
\ell(p)\nu(k)\ell^\prime(q)\nu^\prime(l)$, $M_{T2}$ is given by
 \begin{equation} M_{T2} \equiv
\min_{\mathbf{k}_T+\mathbf{l}_T=\psl_T} \left[\max\left\{
M_T^{(1)},\,M_T^{(2)}\right\}\right],
\end{equation}
where $M_T^{(1)}$ and $M_T^{(2)}$  are the transverse masses of the
two leptonically decaying $W$ bosons:
\begin{eqnarray}
\left(M_T^{(1)}\right)^2
= 2\left(|\mathbf{p}_T||\mathbf{k}_T|
-\mathbf{p}_T\cdot\mathbf{k}_T\right), \quad
\left(M_T^{(2)}\right)^2 = 2\left(|\mathbf{q}_T||\mathbf{l}_T|
-\mathbf{q}_T\cdot\mathbf{l}_T\right) .
\end{eqnarray}
This $M_{T2}$ has an endpoint at $m_H/2$ when $m_H\leq
2m_W$~\cite{Choi:2009hn}, suggesting that $M_{T2}$ also might be
useful for the Higgs boson mass measurement.

In fact, with $M_{T2}$ and additional on-shell constraints, one can
approximately reconstruct the invisible particle 4-momenta in each
event~\cite{Cho:2008tj}. This method of ``$M_{T2}$-assisted-on-shell
(MAOS) reconstruction'' of invisible particle momenta has been
applied to the Higgs boson event $H\rightarrow WW\rightarrow
\ell(p)\nu(k)\ell^\prime(q)\nu^\prime(l)$~\cite{Choi:2009hn} in
order to examine the invariant mass variable
\begin{equation}
\left(m_H^{\rm maos}\right)^2 \equiv \left(p+q+k^{\rm maos}+l^{\rm
maos}\right)^2,
\end{equation}
where $k^{\rm maos}$ and $l^{\rm maos}$ are the reconstructed
neutrino 4-momenta. It was then argued that the distribution of this
MAOS Higgs mass exhibits a peak at the true Higgs boson mass.
Furthermore, the reconstructed  MAOS momenta $k^{\rm maos}$ and
$l^{\rm maos}$ are closer to the true neutrino momenta for the
events near the upper endpoint of $M_{T2}$. As a result, the
$m_H^{\rm maos}$ distribution can exhibit even a narrow resonance
peak at $m_H$ when a strong $M_{T2}$ cut is employed. These
observations suggest that $m_H^{\rm maos}$ might provide a powerful
tool to probe the Higgs boson mass through dileptonic $W$ boson
decays.

%

Since it has been claimed that 
various kinematic variables 
are useful for the Higgs boson mass measurement, it would be
instructive  to perform a comparative study of the efficiencies of
those variables in the LHC environment.
In this paper, we wish
to perform such a comparative study for $M_T^{\rm true}$, $M_{T2}$,
and $m_H^{\rm maos}$, and explore the possibility to improve the
efficiency of the analysis. This paper is organized as follows.
%
In Sec.~\ref{sec:theo}, we discuss 
the distinctive features of $M_T^{\rm true}$, $M_{T2}$, and $m_H^{\rm
maos}$ for dileptonic $W$ boson decays, mainly focusing on the
possible correlations between these three kinematic variables. 
From the discussion in
Sec.~\ref{sec:theo}, it is 
evident that $M_{T2}$ is less
efficient than $M_T^{\rm true}$ as a main observable for the Higgs
boson search. However, $M_{T2}$ might be 
a useful cut
variable in the analysis using $M_T^{\rm true}$ or $m_H^{\rm maos}$
as the main observable.
As for the comparison of $M_T^{\rm true}$ and $m_H^{\rm maos}$, it
is not obvious which one is more efficient since each  variable has
its own virtue and weak point.  We thus perform in
Secs.~\ref{sec:GF} and \ref{sec:VBF} a detailed
comparative study of  $M_T^{\rm true}$ and $m_H^{\rm maos}$.
 Our
results show that 
$m_H^{\rm maos}$ and $M_T^{\rm true}$ will eventually
have a similar efficiency for both the GGF and VBF Higgs
productions at the LHC. 
In this study, we also examine
the possible role of $M_{T2}$ as a  cut variable, and find that
the Higgs signal distribution for the GGF channel becomes narrower
under the $M_{T2}$ cut, while the background distribution becomes
flatter. Although an estimate of any systematic uncertainty is
beyond the scope of this paper,  such behavior of the
signal/background distributions under the $M_{T2}$ cut indicates
that one might be able to reduce the systematic errors in the GGF
channel with an appropriate $M_{T2}$ cut.
We summarize our conclusions in Sec.~\ref{sec:concl}.

\section{Distinctive features of the transverse mass variables}
\label{sec:theo}

Unlike the Higgs boson decay into two photons or two tau
leptons\footnote{ In the di-tau channel, each tau decays to hadron
(or lepton) + a neutrino(s). Although the final state neutrinos  make
a full reconstruction of the event impossible, it is known that the
di-tau invariant mass can be determined well in collinear
approximation~\cite{Ellis:1987xu}.}, the Higgs boson mass cannot be
reconstructed  in the event $H \to WW^{(\ast)} \to \ell(p)\nu(k) \,
\ell^\prime(q)\nu^{\prime}(l)$ due to the missing neutrinos.
One then often considers the transverse mass of the $W$ pair, $M_T(WW)$,
defined in Eq.~(\ref{eq:mtww}), as the main observable to probe the
Higgs signal
event~\cite{Aad:2009wy,Asai:2004ws,Rainwater:1999sd,Barr:2009mx}. If
one could determine the correct value of $m_{\nu\nu^\prime}$
event by event, the resulting distribution of $M_T(WW)$ would have
an upper endpoint at $m_H$. However, $m_{\nu\nu^\prime}$ is not
available, in general, and there have been two proposals to fix the
unknown $m_{\nu\nu^\prime}$.
The first one is to choose $m_{\nu\nu^\prime} =
m_{\ell\ell^\prime}$~
\cite{Aad:2009wy,Asai:2004ws,Rainwater:1999sd}, and consider the
distribution of
\begin{eqnarray}
M_T^{\rm approx}\equiv
\left.M_T(WW)\right|_{m_{\nu\nu^\prime}=m_{\ell\ell^\prime}}.\end{eqnarray}
This has been motivated by the observation  that, for a Higgs
boson mass close to $2m_W$, the $W$ bosons are produced at
near-threshold and almost at rest in their center-of-mass frame, for
which $m_{\nu\nu^\prime}^{\rm true}=m_{\ell\ell^\prime}$, where
$m_{\nu\nu^\prime}^{\rm true}$ is the true value of
$m_{\nu\nu^\prime}$
 in the event.
 Although
this threshold approximation is not necessarily
suitable
for generic events and therefore  $M_T^{\rm approx}$ is not strictly
bounded from above by $m_H$, still the shape and range of its
distribution can provide information on the Higgs boson mass.
The second proposal~\cite{Barr:2009mx} is deduced from the simple
relation
\begin{equation}
M_T^{\rm true}\equiv M_T(WW) |_{m_{\nu\nu^\prime}=0}\, \leq\,
M_T(WW) |_{m_{\nu\nu^\prime}=m_{\nu\nu^\prime}^{\rm true}} \,\leq\,
m_H . \label{eq:trans_rel}
\end{equation}
%
It was reported that one can obtain a more accurate Higgs boson mass
by using $M_T^{\rm true}$
rather than using $M_T^{\rm approx}$
\cite{Barr:2009mx,DPinNath:2010zj}.

For the event  $H\rightarrow WW^{(*)}\rightarrow
\ell(p)\nu(k)\ell^\prime(q)\nu^\prime(l)$, one may exploit the event
variable $M_{T2}$ which has been
designed for a general situation with two invisible particles (with
the same mass) in the final state~\cite{refs:mt2}. The variable is
defined as
\begin{equation}
M_{T2} \equiv \min_{\mathbf{k}_T+\mathbf{l}_T=\psl_T}
\left[\max\left\{ M_T^{(1)},\,M_T^{(2)}\right\}\right],
\end{equation}
where $M_T^{(1)}$ and $M_T^{(2)}$  are the transverse masses of the
two leptonically decaying $W$ bosons:
\begin{eqnarray}
\left(M_T^{(1)}\right)^2
= 2\left(|\mathbf{p}_T||\mathbf{k}_T|
-\mathbf{p}_T\cdot\mathbf{k}_T\right), \quad
\left(M_T^{(2)}\right)^2 = 2\left(|\mathbf{q}_T||\mathbf{l}_T|
-\mathbf{q}_T\cdot\mathbf{l}_T\right) .
\end{eqnarray}
For a specific set of events for which the Higgs boson has vanishing
transverse momentum, $\mathbf{\psl}_T=-\mathbf{p}_T-\mathbf{q}_T$,
one can use the following analytic expression of $M_{T2}$:
\begin{eqnarray}
  M_{T2}^2 = 2\left(|\mathbf{p}_T||\mathbf{q}_T|+
  \mathbf{p}_T\cdot\mathbf{q}_T\right)
  \end{eqnarray}
to find the upper bound
\begin{eqnarray}
  M_{T2}^2\,
  \leq\,
  \left(|\mathbf{p}_T| + |\mathbf{q}_T|\right)^2
  \,\leq\, \frac{m_H^2}{4}.
  \label{eq:mt2bound}
\end{eqnarray}
Although the above relation is derived for the specific event set
with vanishing Higgs transverse momentum, it is likely that the same
upper bound
applies  for generic Higgs boson events with nonvanishing Higgs
boson transverse momentum. We found through numerical analysis that
it is indeed the case  and the inequality $M_{T2}\leq m_H/2$ remains
true for generic dileptonic Higgs boson events $H\rightarrow
WW^{(*)}\rightarrow \ell\nu\ell^\prime\nu^\prime$.
[See  the later discussion of the inequality between
$M_{T2}$ and $M_T^{\rm true}$ in Eq. (\ref{inequalities}).]
On the other hand, in the case that both $W$ bosons are on shell,
$M_{T2}$ is bounded by $m_W$. Combining this with (\ref{eq:mt2bound}),
one may find that the
$M_{T2}$ of the Higgs boson decay is bounded by~\cite{Choi:2009hn}
\begin{equation}
M_{T2}^{\rm max} = \left\{
\begin{array}{ll}
m_H / 2 & \mbox{for}\,\,\, m_H \leq 2m_W \\
m_W & \mbox{for}\,\,\,m_H \geq 2m_W ,
\end{array}
\right .
\label{eq:mt2max_signal}
\end{equation}
implying that $M_{T2}$ can provide information on the Higgs boson
mass through its endpoint.
%

Apart from $M_T^{\rm approx}$, $M_T^{\rm true}$, and $M_{T2}$ in
dileptonic $W$ boson decays, there has been a proposal to
reconstruct an invariant mass by using the
MAOS reconstruction of the two neutrino
momenta~\cite{Choi:2009hn}.
In MAOS reconstruction,
the transverse momenta are defined as the ones that determine the
value of $M_{T2}$, i.e.
\begin{eqnarray} 
M_{T2}({\bf p}_T, {\bf q}_T,\psl_T)\,=\,M_T^{(1)}({\bf p}_T,{\bf k}_T^{\rm
maos})\,=\,M_T^{(2)}({\bf q}_T,{\bf l}_T^{\rm maos}=\psl_T-{\bf
k}_T^{\rm maos}), \label{maos_tr}
\end{eqnarray}
while the longitudinal and energy components are
obtained from the following constraints\footnote{Note that one could
use $m_W$ instead of $M_{T2}$ in the second constraints if both of
the $W$ bosons are on shell. Our scheme to use $M_{T2}$ can be
applicable irrespective of whether the $W$ bosons are on shell or
off shell.}:
\begin{equation}
\left(k^{\rm maos}\right)^2=\left(l^{\rm maos}\right)^2=0, \quad
\left(p+k^{\rm maos}\right)^2=\left(q+l^{\rm maos}\right)^2=M_{T2}^2
.\label{maos_lo}
\end{equation}
It was noticed that the MAOS 4-momenta determined as the solutions
of (\ref{maos_tr}) and (\ref{maos_lo}) provide a reasonable
approximation to the true neutrino 4-momenta, and they become closer
to the true momenta for the near-endpoint events in the $M_{T2}$
distribution~\cite{Cho:2008tj,Choi:2009hn}. Then the distribution of
the following invariant mass ($\equiv$ MAOS Higgs boson mass),
\begin{equation}
\left(m_H^{\rm maos}\right)^2 \equiv \left(p+q+k^{\rm maos}+l^{\rm
maos}\right)^2\,
\end{equation}
exhibits a peak at the true Higgs boson mass,
and
the peak shape becomes narrower if one imposes an event cut
selecting the events near the endpoint of $M_{T2}$. Although such an
$M_{T2}$ cut can 
cause the $m_H^{\rm maos}$ distribution to have  a
resonance peak at $m_H$ and also might eliminate some of the
backgrounds, it can also sacrifice the number of the Higgs signal, thus
worsening the statistical significance of the mass measurement.
Therefore for a given luminosity 
a careful analysis is required to
see if the efficiency of the Higgs mass measurement can be improved
by a proper $M_{T2}$ cut.

To proceed, let us consider a specific set of events for which the
$W$ boson pair has a small transverse momentum: ${\bf
p}_T^{WW}\approx 0$. Although it does not cover the full event set,
this subset of events reveal some of the essential features of the
kinematic variables under discussion. For those events with small
${\bf p}_T^{WW}$, we have
\begin{eqnarray}
\psl_T= -({\bf p}_T+{\bf q}_T),\nonumber\end{eqnarray} and then the
following simple  expressions of $M_{T2}$ and MAOS momenta are
available~\cite{Cho:2008tj,Choi:2009hn}:
\begin{eqnarray}
 &&{\bf k}_T^{\rm maos} = -{\bf q}_T\,,  \quad k_L^{\rm
maos}=\frac{|{\bf k}_T^{\rm maos}|}{|{\bf p}_T|}\,p_L\,,
\nonumber \\
&&{\bf l}_T^{\rm maos}\, = -{\bf p}_T\,, \quad l_L^{\rm
maos}=\frac{|{\bf l}_T^{\rm maos}|}{|{\bf q}_T|}\,q_L\,, \nonumber
\\
&&M_{T2}^2=2 \left(|{\bf p}_T| |{\bf q}_T|+ {\bf p}_T \cdot {\bf
q}_T\right)\,. \label{eq:maosmomenta}
\end{eqnarray}
With these expressions of $M_{T2}$ and MAOS momenta, it is
straightforward to find
\begin{eqnarray}
\left(m_H^{\rm maos}\right)^2&=& \left(2+\left|\frac{{\bf p}_T}{{\bf
q_T}}\right|+\left|\frac{{\bf q}_T}{{\bf
p_T}}\right|\right)\left(2\left(|{\bf p}_T| |{\bf q}_T|+ {\bf p}_T
\cdot {\bf q}_T\right)+ m_{\ell\ell^\prime}^2\right),\nonumber
\\
M_T^{\rm true}&=& \left|{\bf p}_T+{\bf q}_T\right|+\sqrt{\left|{\bf
p}_T+{\bf q}_T\right|^2+m_{\ell\ell^\prime}^2}.
\end{eqnarray}
Since
\begin{eqnarray}
M_{T2}^2&=&2 \left(|{\bf p}_T| |{\bf q}_T|+ {\bf p}_T \cdot {\bf
q}_T\right)\,\leq\,\left|{\bf p}_T+{\bf q}_T\right|^2, \nonumber \\
m_{\ell\ell^\prime}^2&=& 2\left(\left|{\bf p}\right|\left|{\bf
q}\right|-{\bf p}\cdot{\bf q}\right)\,\geq \, 0,
\end{eqnarray}
both $m_H^{\rm maos}$ and $M_T^{\rm true}$ are bounded below by $2
M_{T2}$, and therefore
\begin{eqnarray}
\label{inequalities} 2M_{T2}\,\leq\, M_T^{\rm true} \,\leq \,
m_H\,,\quad 2M_{T2}\, \leq\, m_H^{\rm maos} \,.  \label{eq:m.bounds}
\end{eqnarray}
%

Although the above inequalities between $M_{T2}$, $M_T^{\rm true}$ 
and $m_H^{\rm maos}$ could be derived analytically
only for the events with ${\bf p}_T^{WW}=0$, we find numerically
that they  hold true  also for the events with 
${\bf p}_T^{WW}\neq 0$.
  In Fig.~\ref{fig:mh_0} (left panel), we show the distributions of
these three kinematic variables for the true Higgs mass $m_H=160$
GeV. Here we use the Higgs boson events produced by
the 
GGF 
process at the LHC, while incorporating
the detector effects with the fast detector simulation program {\tt
PGS4}~\cite{pgs}.
%
Figure ~\ref{fig:mh_0} (left panel) shows, first of all, that both
$M_T^{\rm true}$ and $2M_{T2}$ have a common  upper endpoint at
$m_H=160$ GeV, as we have anticipated in the above discussion.
However, the $M_T^{\rm true}$ distribution is significantly narrower
than the $2M_{T2}$ distribution, as implied by the first inequality
in Eq.~(\ref{eq:m.bounds}). This indicates that $M_T^{\rm true}$ is
more sensitive to $m_H$ than $M_{T2}$, and thus is a more
efficient variable to determine $m_H$. If one compares $M_T^{\rm
true}$ to $m_H^{\rm maos}$, Fig.~\ref{fig:mh_0} (left panel) shows
that $m_H^{\rm maos}$ has a peak at $m_H$, while  $M_T^{\rm true}$
has an endpoint at $m_H$. With this feature, one might expect that
$m_H^{\rm maos}$ is more sensitive to $m_H$ than $M_T^{\rm true}$.
However the $M_T^{\rm true}$ distribution around $m_H$ is
significantly narrower than the $m_H^{\rm maos}$ distribution, and
thus one needs a detailed analysis to see which of $M_T^{\rm true}$
and $m_H^{\rm maos}$ is more efficient for the determination of
$m_H$. In the next two sections, we perform such an analysis for both
the GGF Higgs production (Sec.~\ref{sec:GF}) and the VBF Higgs
production (Sec.~\ref{sec:VBF}) at the LHC.

\section{Measuring the Higgs boson mass in GGF}
\label{sec:GF}
In this section, we consider the SM Higgs boson production at the
LHC through the 
GGF 
process and its decay into two
leptonically decaying $W$ bosons.
To investigate the experimental performance of $m_H^{\rm maos}$ and
$M_T^{\rm true}$, we use the {\tt PYTHIA6.4}
event generator~\cite{Sjostrand:2000wi}
at the proton-proton center-of-mass energy of 14 TeV.
The generated events have been
further processed through the fast detector simulation program {\tt
PGS4}~\cite{pgs} to incorporate the detector effects with reasonable
efficiencies and fake rates.
The dominant background comes from the continuum $q\bar{q}\,,\,gg
\to WW \to l\nu l^\prime\nu^\prime$ process, and we also include the
$t\bar{t}$ background in which the two top quarks decay into a pair
of $W$ bosons and two $b$ jets.
For the details of the MC event samples of the SM Higgs boson signal
and the two main backgrounds, and  also the employed  event
selection cuts, we refer to Ref.~\cite{Choi:2009hn}.

\begin{figure}[t!]
\begin{center}
\epsfig{figure=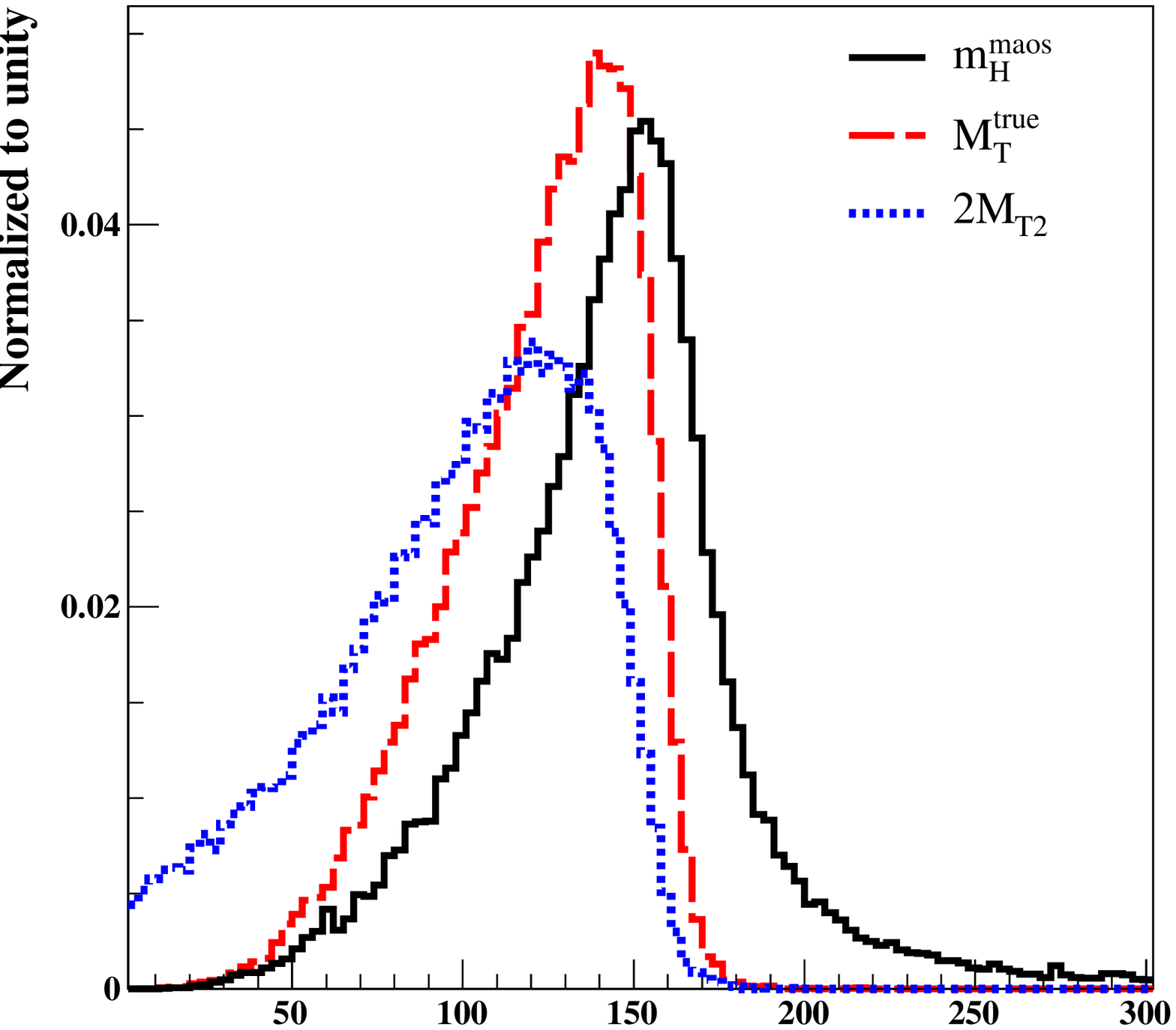,height=8.0cm,width=8.0cm}
\epsfig{figure=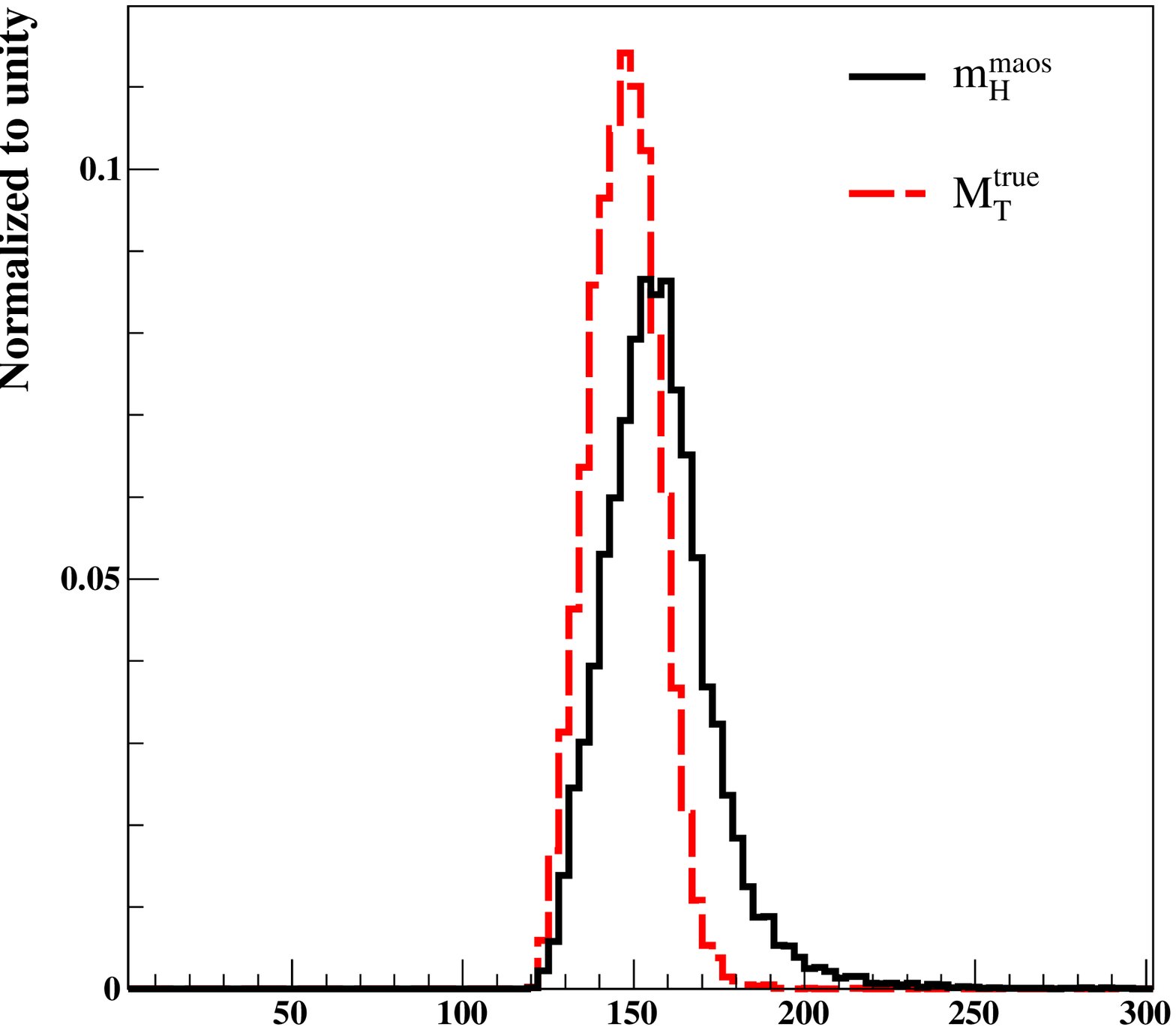,height=8.0cm,width=8.0cm}
\end{center}
\caption{ The signal $m_H^{\rm maos}$ (black solid line) and $M_T^{\rm
true}$
  (red dashed line) distributions at the detector-level
without (left panel) and with (right panel) the $M_{T2}$ cut, $M_{T2}>60$ GeV.
  The Higgs boson mass is taken to be 160 GeV, and
we use the same event samples as in Ref.~\cite{Choi:2009hn}. }
\label{fig:mh_0}
\end{figure}
\begin{figure}[t!]
\begin{center}
\epsfig{figure=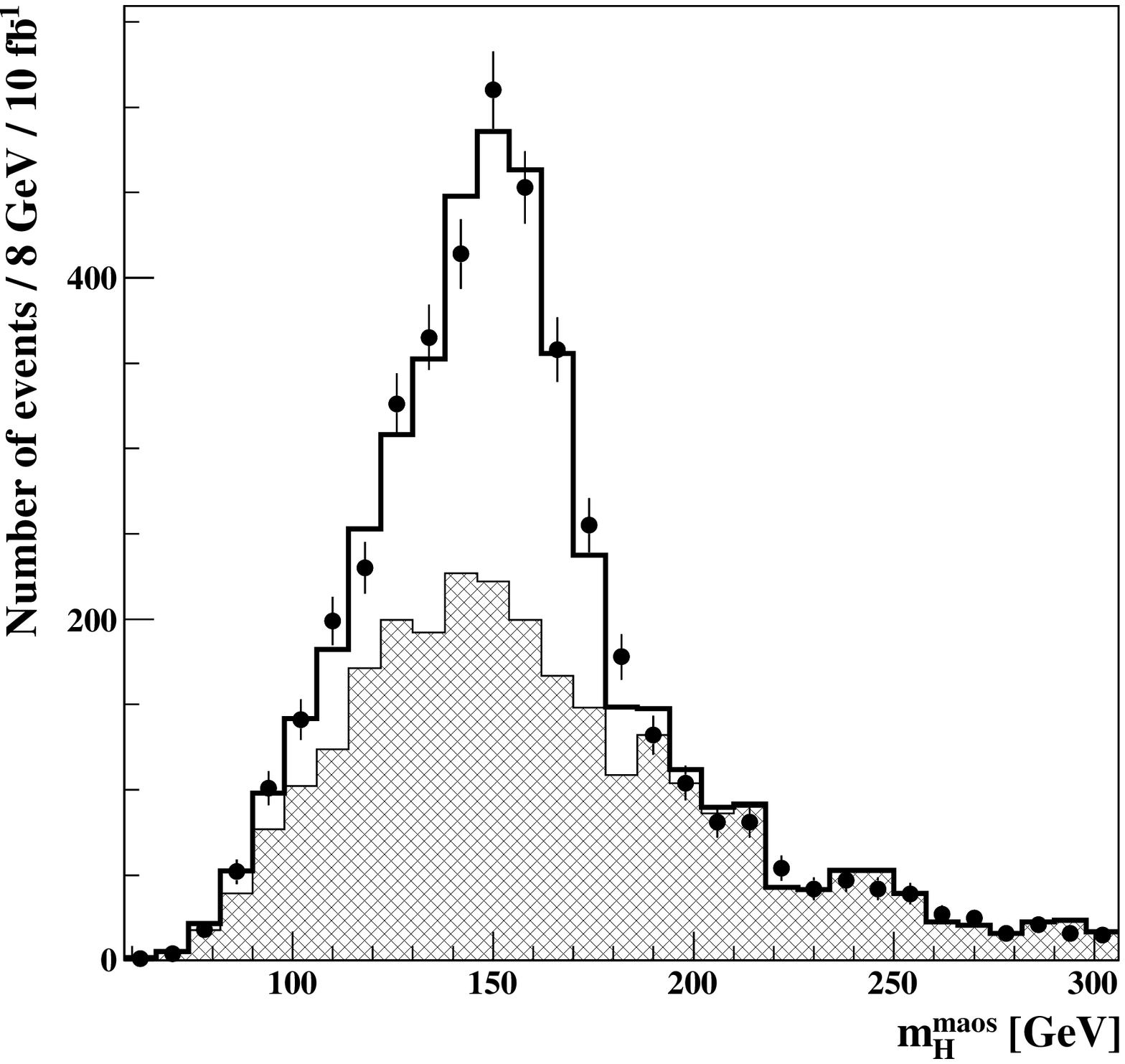,height=8.0cm,width=8.0cm}
\epsfig{figure=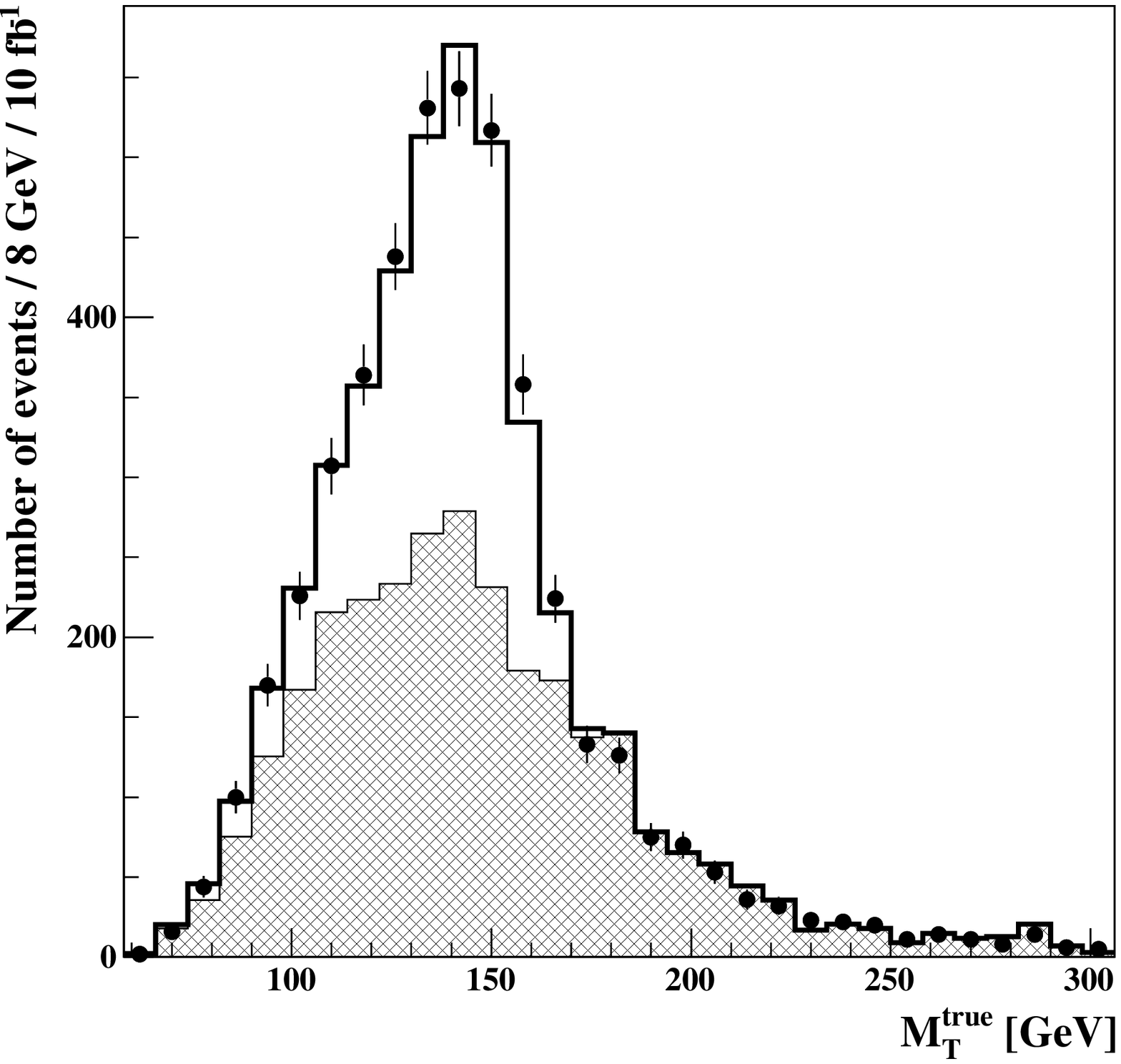,height=8.0cm,width=8.0cm}
\epsfig{figure=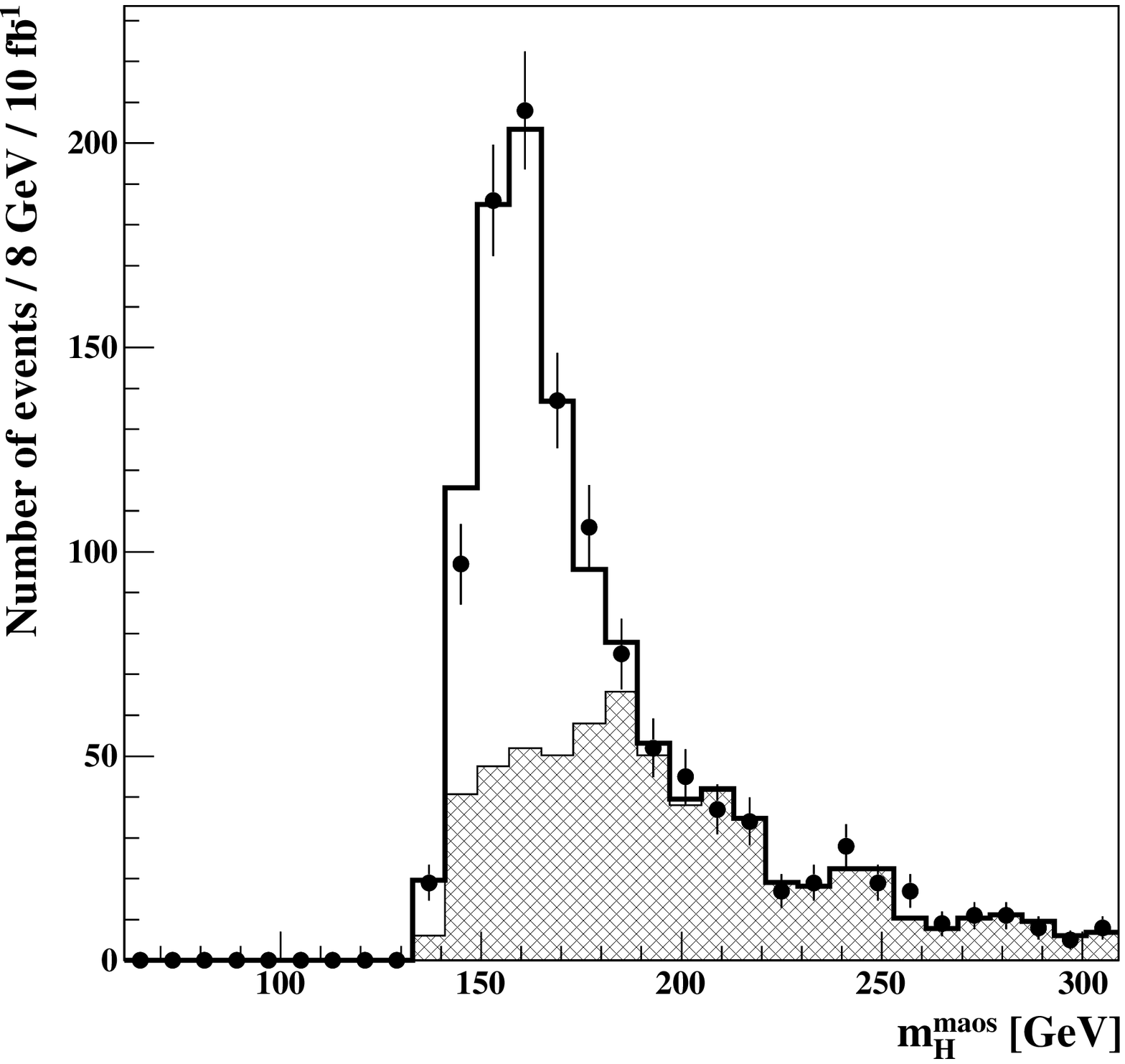,height=8.0cm,width=8.0cm}
\epsfig{figure=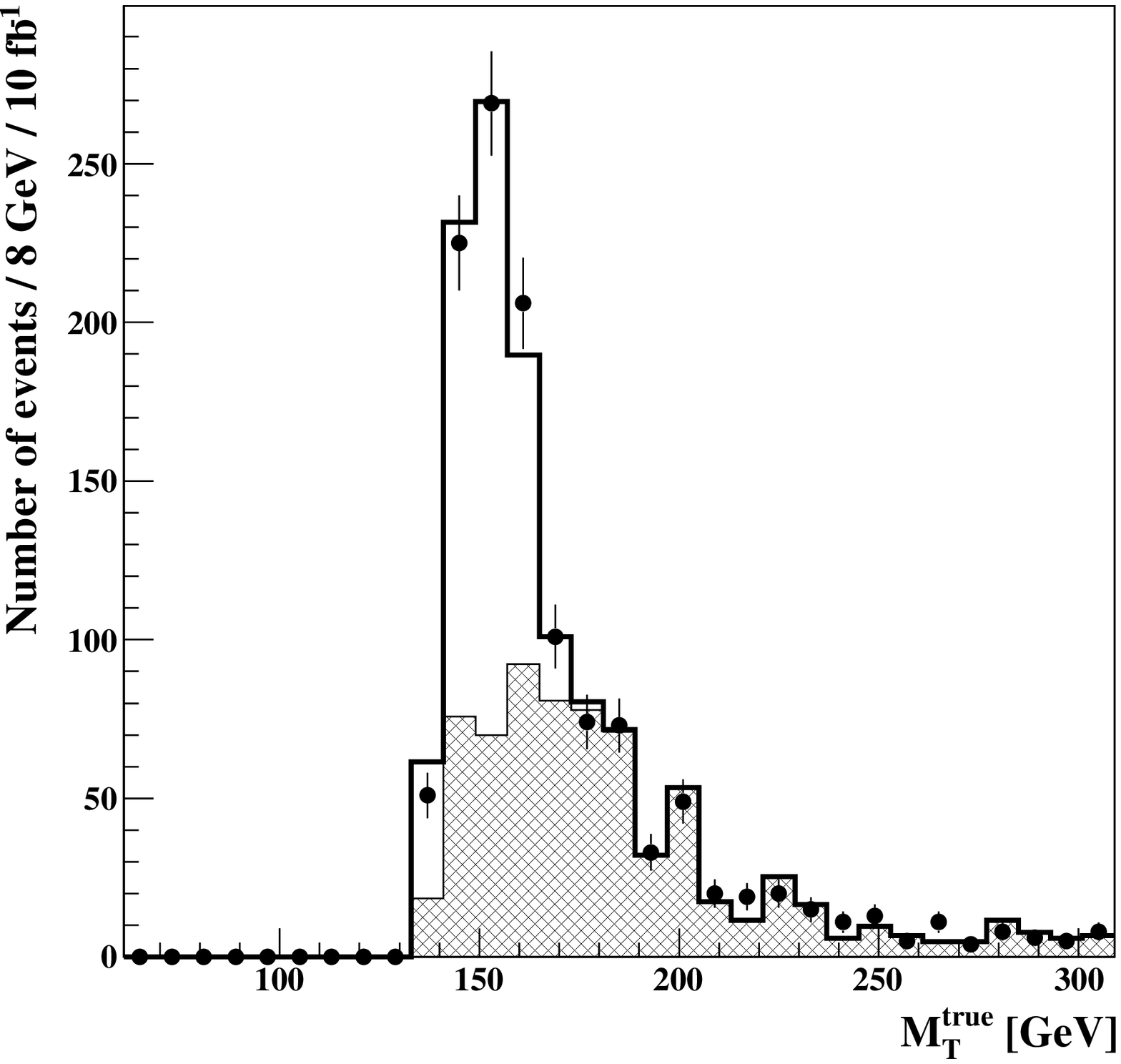,height=8.0cm,width=8.0cm}
\end{center}
\caption{ The $m_H^{\rm maos}$ (left frames) and the $M_T^{\rm
true}$ (right frames) distributions at 10 fb$^{-1}$
luminosity. In each frame, the shaded region represents the
backgrounds ($t\bar{t}$ and $WW$) and the event selection cuts
without (upper frames) and with (lower frames) the $M_{T2}$ cut
($M_{T2}>66$ GeV) 
imposed. The Higgs boson mass is taken
to be 160 GeV, and we again refer to Ref.~\cite{Choi:2009hn} for the
details. } \label{fig:mh_1}
\end{figure}
\begin{figure}[t!]
\begin{center}
\epsfig{figure=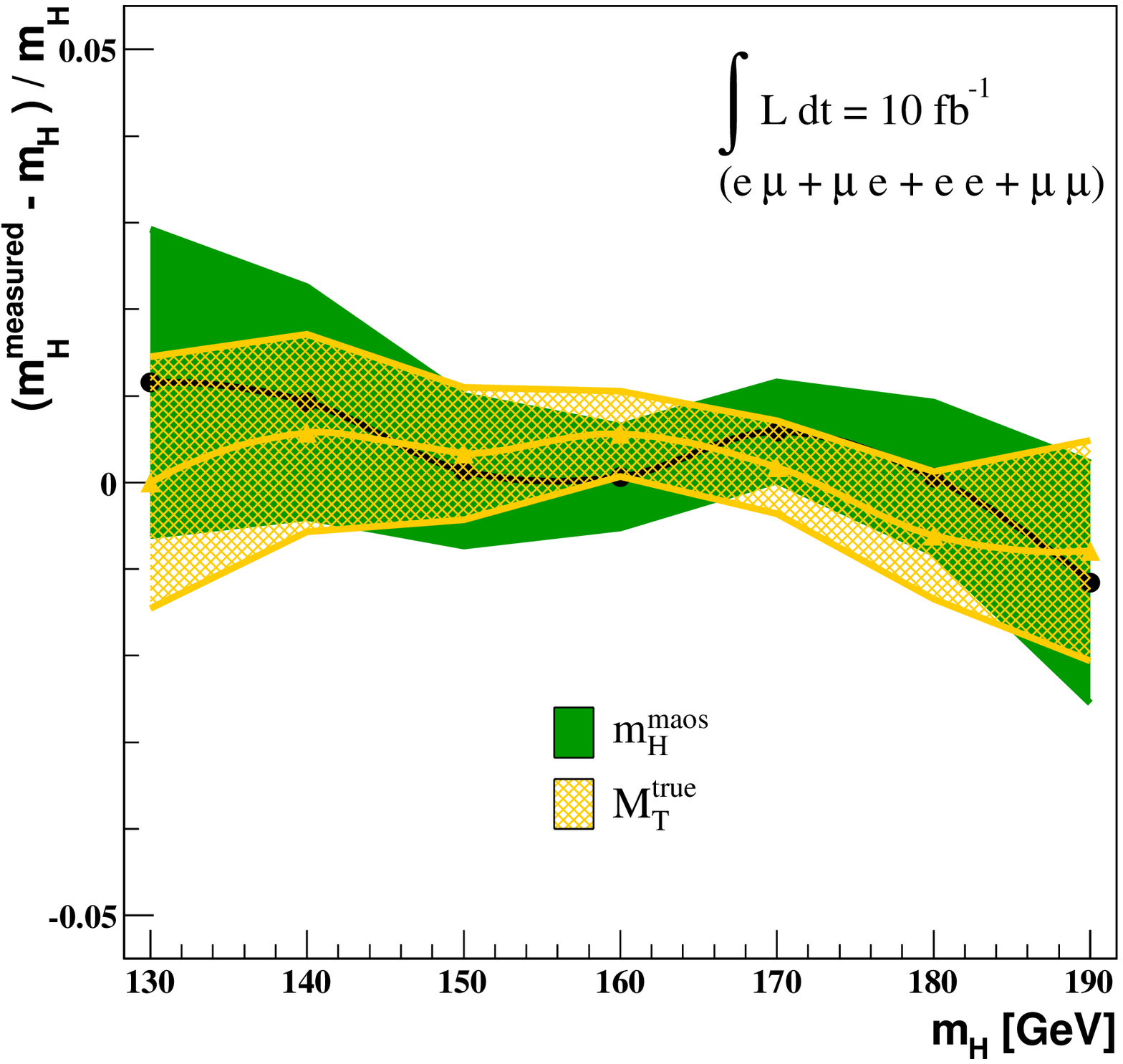,height=8.0cm,width=8.0cm}
\epsfig{figure=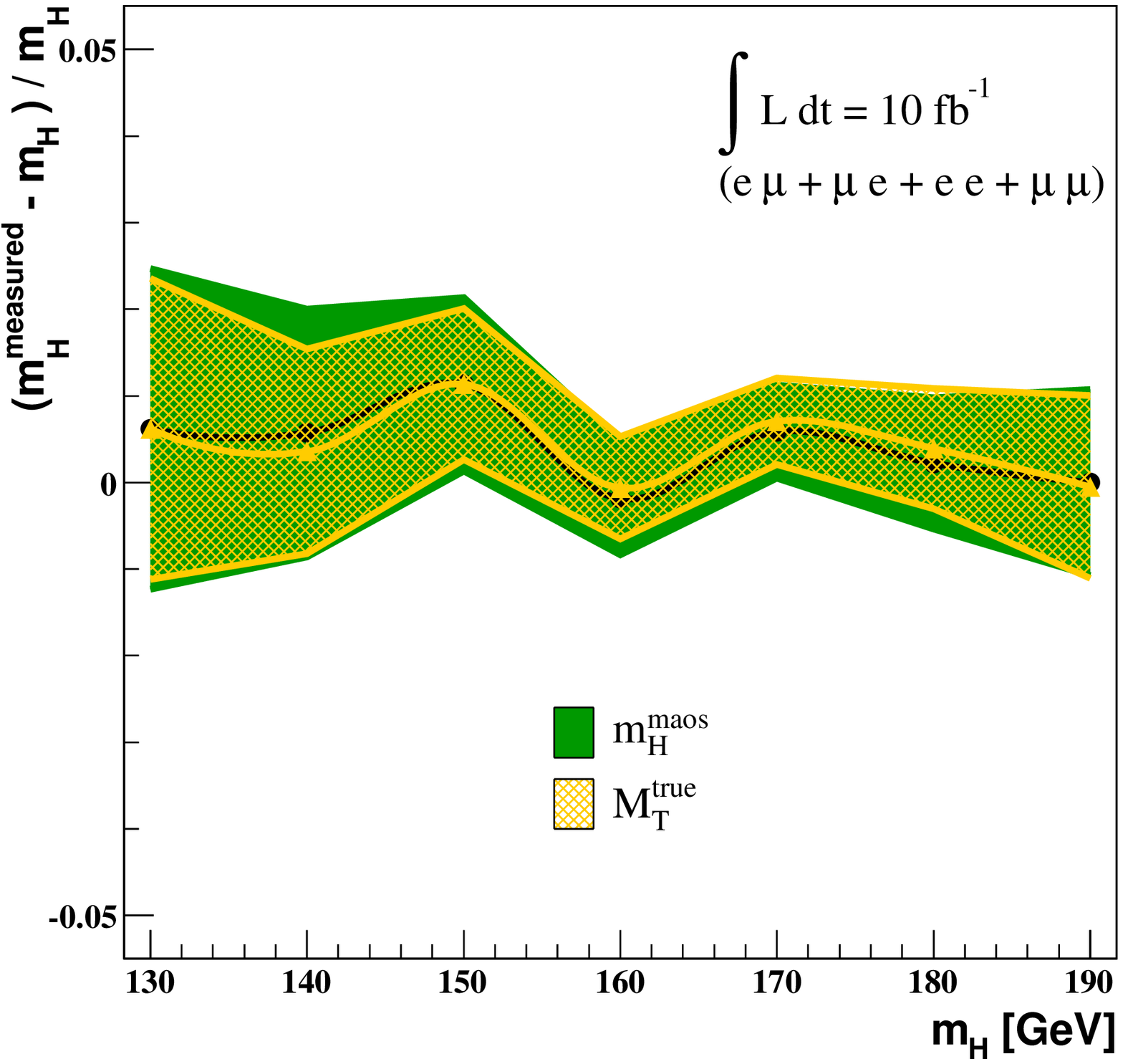,height=8.0cm,width=8.0cm}
\end{center}
\caption{The bands showing the 1-$\sigma$ deviation
of the statistical error for the Higgs boson mass determined by the
$m_H^{\rm maos}$ and $M_T^{\rm true}$ distributions in the GGF
process $gg\to H \to WW^{(\ast)} \to l\nu \, l^\prime\nu^{\,\prime}$.
The dots and lines denote the Higgs boson mass obtained by the
likelihood fit to the distributions. We use the same
event sample and the cut procedure as in Ref.~\cite{Choi:2009hn},
where  the right (left) panel is  the result with (without) an
optimal $M_{T2}$ cut, chosen differently for different values of
$m_H$.} \label{fig:GFcomparison}
\end{figure}

In Fig.~\ref{fig:mh_0}, we show the $m_H^{\rm maos}$ and $M_T^{\rm
true}$ distributions of the signal events  for $m_H=160$ GeV without
(left panel) and with (right panel) the $M_{T2}$ cut: $M_{T2} >
\left(M_{T2}\right)^{\rm cut}= 60$ GeV. We use the same event
samples at the detector level as in Ref.~\cite{Choi:2009hn} but do
not apply the event selection cuts other than the $M_{T2}$ cut.
First we note that the $M_T^{\rm true}$ distribution is bounded
above by $m_H=160$ GeV, while the $m_H^{\rm maos}$ distribution is
peaked around it, as noticed in the discussion of the previous
section. Under an $M_{T2}$ cut, both distributions are bounded from
below by $2\left(M_{T2}\right)^{\rm cut}=120$ GeV with a rather good
approximation, which can be understood by the inequalities in
(\ref{eq:m.bounds}). Incidentally, in the $m_H^{\rm
maos}$ distribution, the region above $m_H$ is less sensitive to the
$M_{T2}$ cut than the region below $m_H$. We also see clearly that
the $M_T^{\rm true}$ distribution is narrower than the $m_H^{\rm
maos}$ distribution as implied by Eq.~(\ref{eq:m.bounds}).
Next, we proceed to perform the template likelihood fitting to the
$M_T^{\rm true}$ and $m_H^{\rm maos}$ distributions. Here, a template
means a simulated distribution with a trial Higgs mass which, in
general, is different from the nominal one used to generate the
data. For each pseudoexperiment distribution with the nominal Higgs
mass $m_H$, 11 templates are generated with the trial Higgs masses
between $m_H-10$ GeV and $m_H+10$ GeV, in steps of 2 GeV. For
example, in each frame of Fig.~\ref{fig:mh_1}, the solid line shows
the template distribution when the trial Higgs mass is the same as
the nominal one. Each template is normalized to the corresponding
pseudoexperiment distribution.
For the definitions of the likelihood between a pseudoexperiment
data distribution and a template, and also for more on the Higgs
masses and 1-$\sigma$ errors obtained by fitting the log likelihood
distributions, we again refer to Ref.~\cite{Choi:2009hn}.

Empirically, the template likelihood fitting, which is used to
estimate the efficiency of the Higgs mass determination in this
work, provides a better result when the width of the distribution is
narrower.
On the other hand, compared to the peak of the distribution which is
generally easier to determine, the endpoints are more vulnerable to
detector smearing, backgrounds and low statistics.
Even though the peak of the $M_T^{\rm true}$ distribution does not
have a strong correlation with the input Higgs mass like that
of the $m_H^{\rm maos}$ distribution, the overall shape has a
definite correlation with the Higgs boson mass and, moreover, its
narrower width might result in a comparable sensitivity to
$m_H^{\rm maos}$. The problem is the shape of the background
distribution which could smear the edge structure of $M_T^{\rm
true}$.
In Fig.~\ref{fig:mh_1}, we have shown the $m_H^{\rm maos}$ (left panels)
and $M_T^{\rm true}$ (right panels) distributions for the Higgs signal and
the backgrounds. We use the same detector-level event samples as in
Ref.~\cite{Choi:2009hn}, and all the event selection cuts are
applied.

Our MC study shows that, generically, the background distributions for
both $M_T^{\rm true}$ and $m_H^{\rm maos}$  become flatter if we
impose a stronger $M_{T2}$ cut. On the other hand, a stronger
$M_{T2}$ cut causes the signal distributions to have a narrower shape
(see Fig.~\ref{fig:mh_0}). This behavior of the signal and background
under the $M_{T2}$ cut suggests that one might be able to improve the
efficiency of the Higgs mass measurement with an appropriate
$M_{T2}$ cut, particularly reduce the systematic uncertainties
associated with various origins such as the detector
energy resolution, fit routines and poorly estimated backgrounds.
As for statistical error, one needs a detailed analysis, as the
improvement due to better shapes of signal and background
distributions can be compensated by the reduced statistics. Indeed,
for an integrated luminosity $\sim\,10$ fb$^{-1}$, we find
(Fig.~\ref{fig:GFcomparison}) that there is no appreciable
improvement of the statistical error gained by the $M_{T2}$ cut,
although the situation might be different at higher luminosity.

In Fig.~\ref{fig:GFcomparison}, we show the Higgs boson masses and
the 1-$\sigma$ (statistical) errors  obtained by the likelihood fit
to the $m_H^{\rm maos}$ and $M_T^{\rm true}$ distributions in the
GGF process $gg\to H \to WW^{(*)} \to l\nu \,
l^\prime\nu^{\,\prime}$.
%
We see that the reconstructed mass is rather close
to the input one. For the sake of comparison, we have not employed
an $M_{T2}$ cut in the left frame since the cut was not included in
the original suggestion of the $M_T^{\rm true}$
variable~\cite{Barr:2009mx}; however, it is employed in the right
frame.
We observe that the efficiency of $m_H^{\rm maos}$ is comparable to
that of $M_T^{\rm true}$ up to systematic errors, which are not
considered in our analysis. In particular, when the $M_{T2}$ cut is
employed, it is hard to say which one shows a better efficiency.
Still, we note that $m_H^{\rm maos}$ could be a better choice when
the distribution is spoiled by some unknown backgrounds beyond the
SM,  since the peak is less vulnerable to unknown backgrounds than
the endpoint.

\section{Measuring the Higgs boson mass in VBF}
\label{sec:VBF}

In this section, we turn to the 
VBF process,
which is the second most important production channel of the SM
Higgs boson at the LHC.
%
The characteristic feature of the process is the existence of
the two forward tagging jets with suppressed hadronic activity
between them and the central Higgs decay products.
Thanks to the exchanges of the colorless vector bosons in the process,
these features lead to a fairly clean environment with
well-isolated signal events in a low background.
Therefore, the VBF process is useful to measure properties
of the Higgs boson in a hadron collider environment.

The same variables introduced in the GGF process could be used to
determine the Higgs boson mass in VBF, since the tagging jets
provide additional information without touching the decay of the
Higgs boson itself.
To investigate the experimental performance of
the MAOS Higgs mass at the LHC through the VBF process,
we have generated MC event samples of the signal $H+2j$ events
and the main $t\bar{t}$ background events by {\tt PYTHIA6.4}
\cite{Sjostrand:2000wi}. The subleading background, i.e.
electroweak $WW+2j$ event showing  a similar characteristic, has
been generated using {\tt MadGraph/MadEvent} \cite{Alwall:2007st}.
The generated events have been further processed with the fast
detector simulation program {\tt PGS4} \cite{pgs}, which
approximates an ATLAS or CMS-like detector. The {\tt PGS4} program
uses a cone algorithm for jet reconstruction, with a default value of
the cone size $\Delta R = 0.5$, where $\Delta R$ is a separation in
the azimuthal angle and pseudorapidity
plane\footnote{The pseudorapidity is defined as
$\eta=-\ln\tan (\theta/2)$ for the angle $\theta$ between the
particle momentum and the beam axis. Note that the pseudorapidity
for a massless particle is equal to the rapidity defined as
$\eta=\frac{1}{2}\ln (E+p_L)/(E-p_L)$.}. We note that the $b$-jet
tagging efficiency is introduced as a function of the jet transverse
energy and pseudorapidity, with a typical value of about $50\,\%$
in the central region for the high energy jets.

Proceeding in a similar way as in Ref.~\cite{Asai:2004ws},
we have imposed the following basic
selection cuts on the Higgs signal and the backgrounds:
\begin{itemize}
\item Preselection cuts
\item $b$-jet veto
\item Tag-jet conditions (forward jet tagging, leptons between jets, central
jet veto)
\item $\gamma^\ast / Z +$ jets, $Z\rightarrow\tau\tau$ rejection
\end{itemize}
Since the final state consists of two hard jets, two charged leptons
and significant missing transverse energy, the preselection cuts
should be the minimal conditions to be imposed. Specifically, we
require at least two jets with $p_T>20$ GeV and $|\eta|<4.8$, and two
isolated, opposite-sign leptons ($e$ or $\mu$) with $p_T>15$ GeV,
and $|\psl_T|>30$ GeV.
The $b$-jet veto is used to exclude the main $t\bar{t}$ background which contains
two $b$ jets.
The tag-jet conditions should reflect
the characteristic features of the VBF process:
two forward tagging jets with
suppressed hadronic activity between them
and central Higgs decay products.
In this work, we define the tagging jets as the two highest
$p_T$ jets in the event
while rejecting the event if they are in the same hemisphere.
Explicitly, we have imposed the following tag-jet conditions.
\begin{itemize}
  \item[$-$] {Forward jet tagging:}
  the two hardest jets are required to be in opposite hemispheres
  ($\eta_{j_1}\times\eta_{j_2} < 0$)
  and to have a pseudorapidity separation  $|\Delta\eta_{j_1\,j_2}|>3$.
  \item[$-$] {Leptons between jets:}
  both leptons are
  required to be between the tag jets in pseudorapidity.
  \item[$-$] {Central jet veto:}
  the event is rejected
  if it contains any
  extra jet (in addition to the two forward tagging jets)
  with $p_T>20$ GeV
  in the pseudorapidity region $|\eta|<3.2$.
\end{itemize}
The last basic selection cut is to reject the Drell-Yan background
involving the variable [see Eq.~(1)]:
\begin{equation} m_T^{ll\nu}\equiv
M_T(WW)|_{m_{ll}=m_{\nu\nu}=0}.\nonumber
\end{equation} Though the Drell-Yan background has been neglected
in this work, we have included the rejection cut to see its effect
on the signal and the other background events under consideration.
For the $Z\to\tau\tau$ rejection, the di-tau invariant mass is
constructed using the collinear approximation, rejecting the events
with $|M_{\tau\tau}-m_Z|<25$ GeV.

\begin{figure}[t!]
  \begin{center}
  \epsfig{file=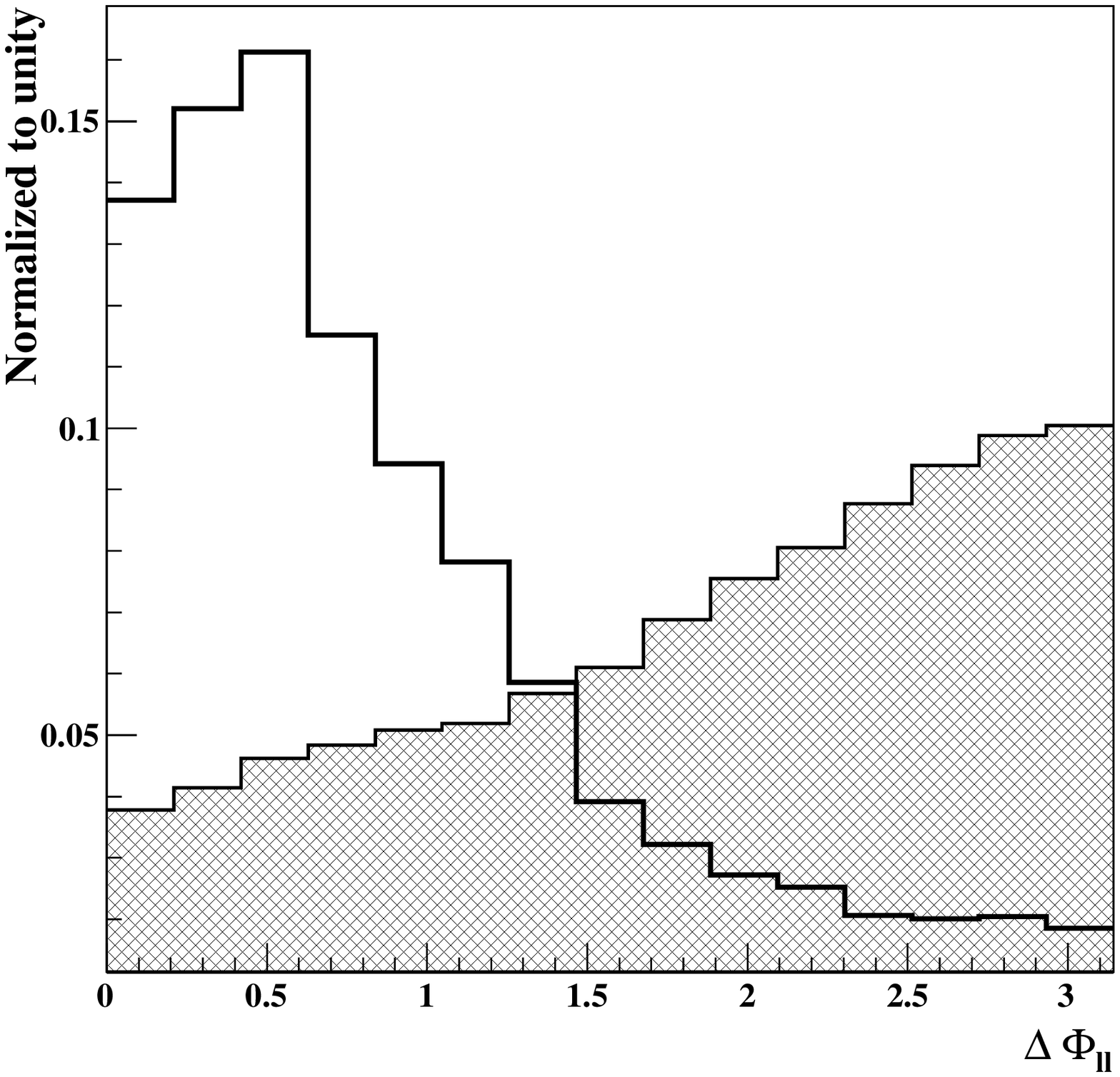,width=8.0cm,height=8.0cm}
  \epsfig{file=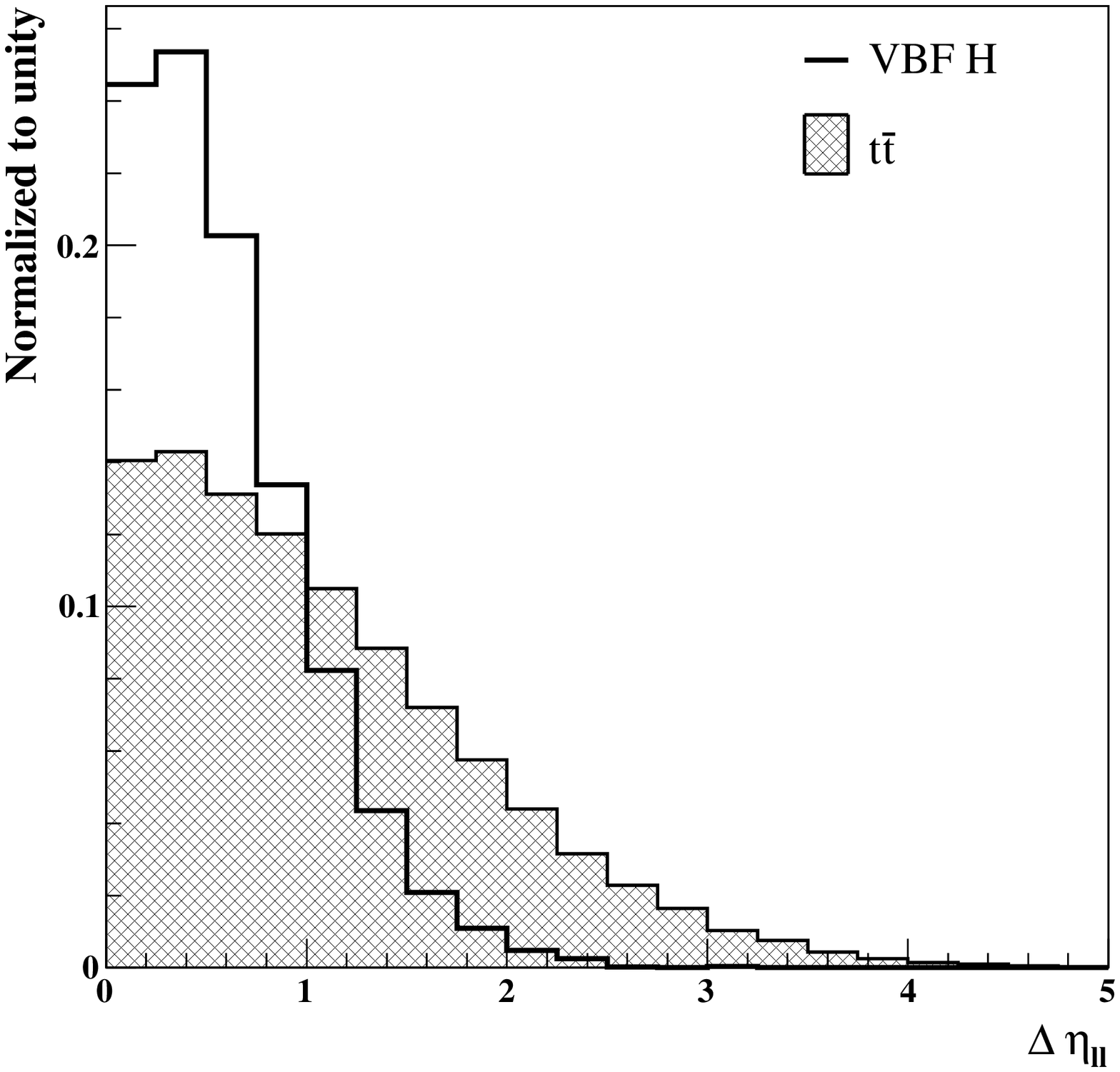,width=8.0cm,height=8.0cm}
  \end{center}
  \caption{Transverse opening angle $\Delta\Phi_{ll}$ (left panel) and
  pseudorapidity gap $\Delta\eta_{ll}$ (right panel)
    between the charged leptons
 in the VBF process
  ($m_H=160$ GeV)
  and the background events.
  The preselection cuts are used for the event selection.
  }
\label{fig:deltall}
\end{figure}
It is well known that the spin-zero nature of the Higgs boson makes two
charged leptons come out in the same direction,
making its opening angle in the Higgs signal smaller
than that in the background; see Fig.~\ref{fig:deltall}.
To incorporate this spin information,
in addition to the basic
selections, we have further required
\begin{equation}
\Delta\Phi_{ll}<\Delta\Phi_{ll}^{\rm cut}\,, \quad
\Delta\eta_{ll}<\Delta\eta_{ll}^{\rm cut}\,,
\end{equation}
where $\Delta\Phi_{ll}$ and $\Delta\eta_{ll}$ are the transverse
opening angle and pseudorapidity gap between the charged leptons.

\begin{figure}[t!]
  \begin{center}
  \epsfig{file=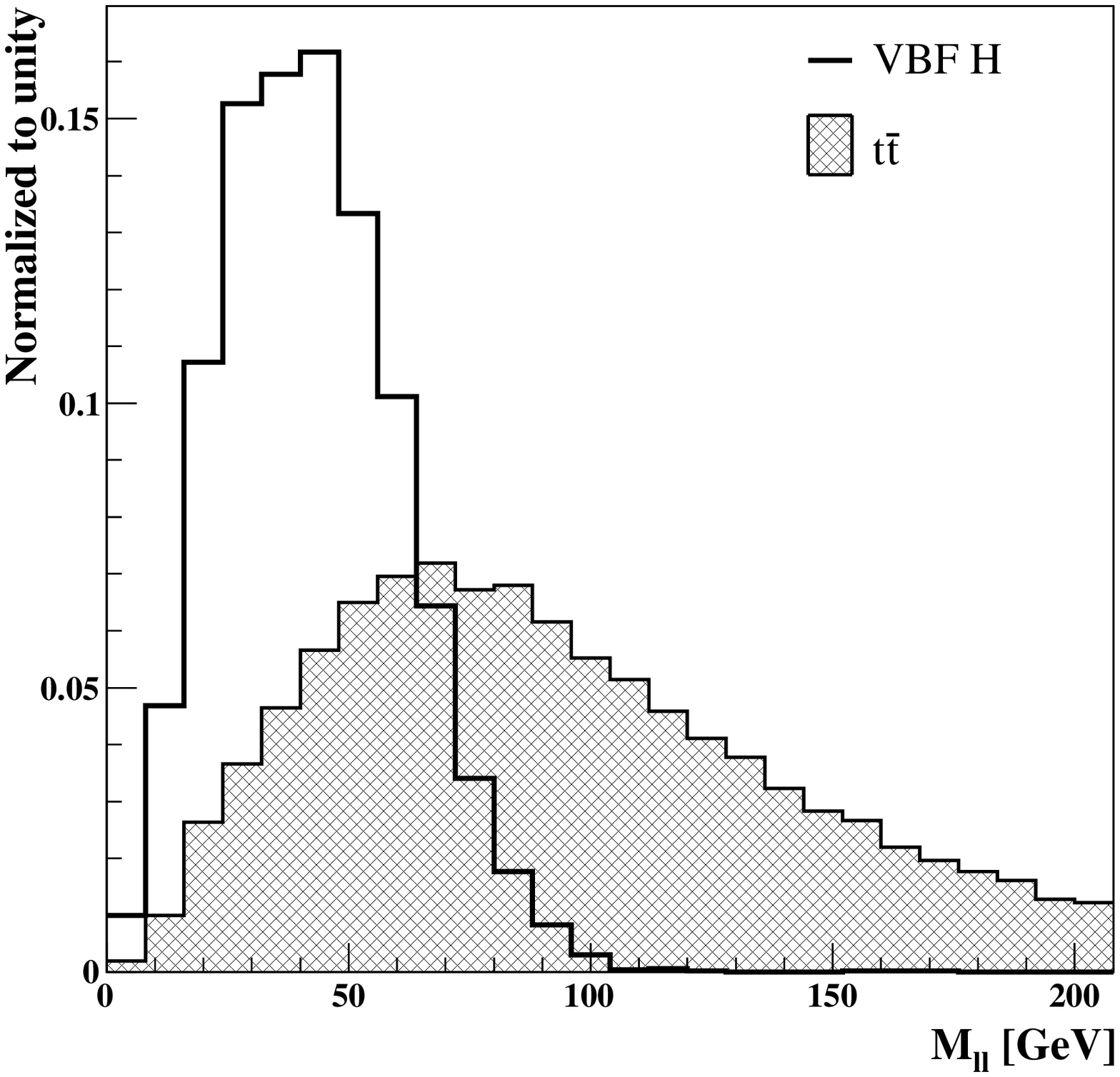,width=8.0cm,height=8.0cm}
  \epsfig{file=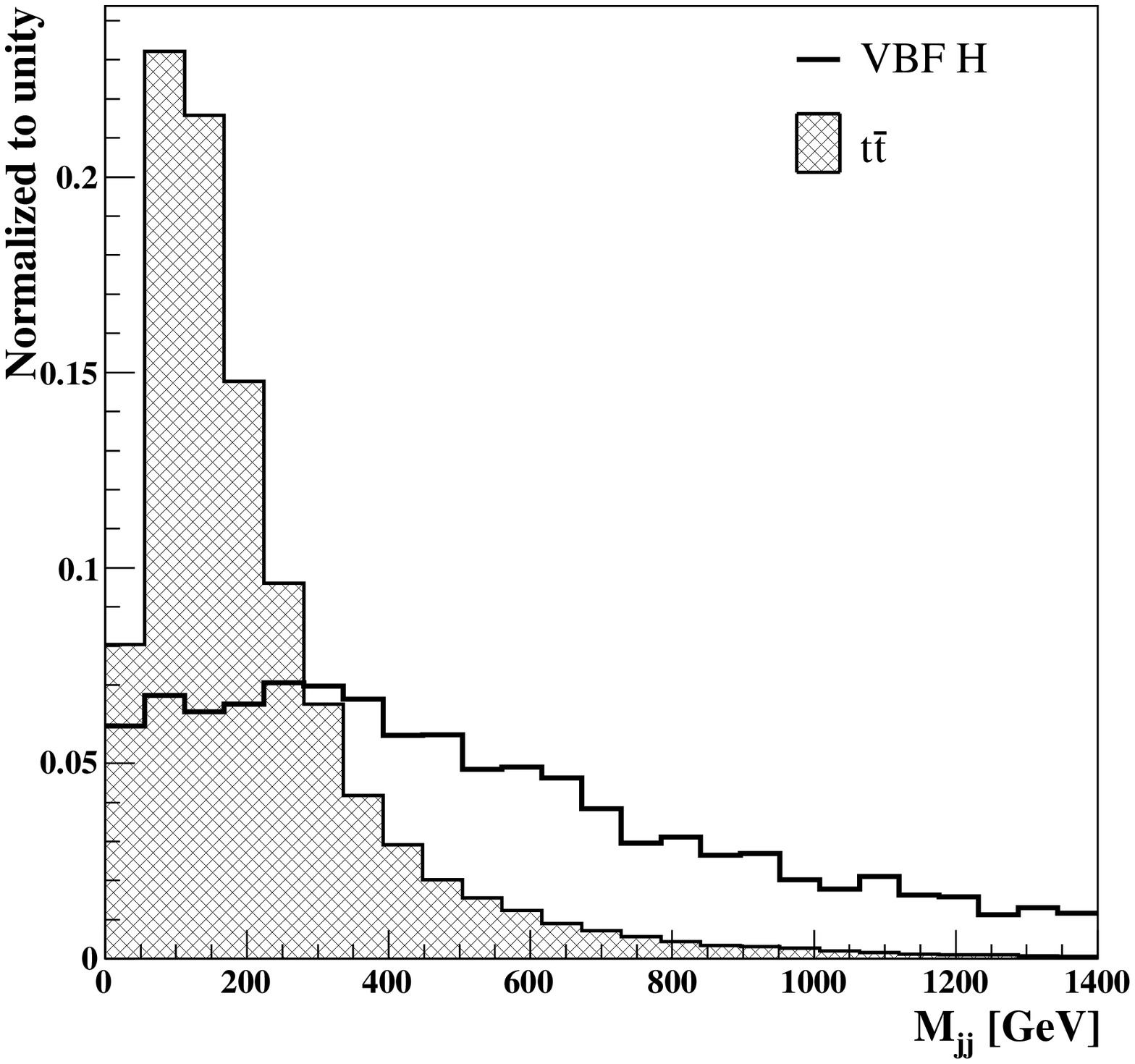,width=8.0cm,height=8.0cm}
  \end{center}
  \caption{The $M_{ll}$ (left panel)
  and $M_{jj}$ (right panel) distributions of the signal
and the $t\bar{t}$ background, after applying the preselection cuts.
$m_H=160$ GeV is taken for the signal.} \label{fig:mllmjj}
\end{figure}

\begin{figure}[t!]
  \begin{center}
  \epsfig{figure=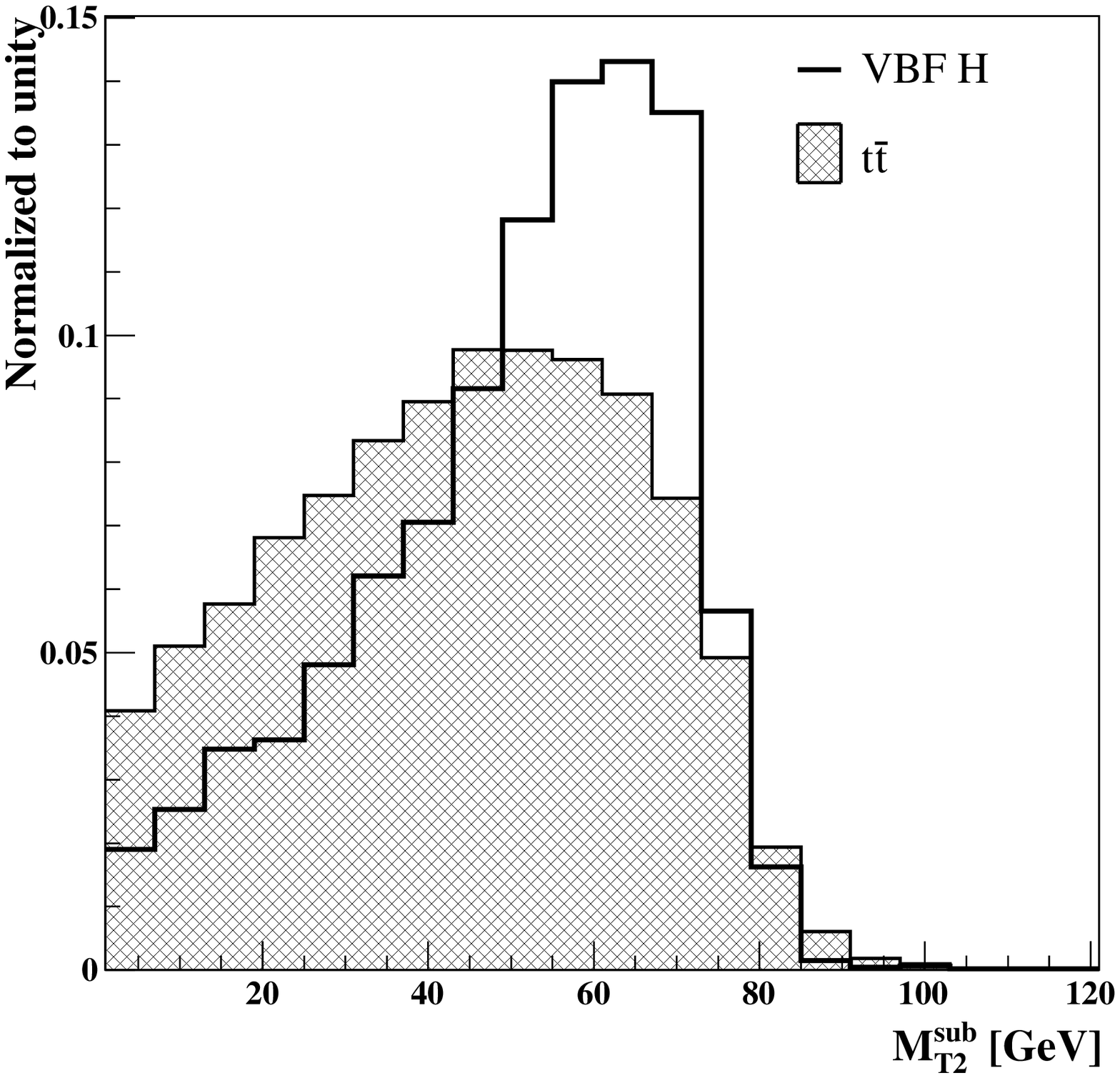, width=8.0cm, height=8.0cm}
  \epsfig{figure=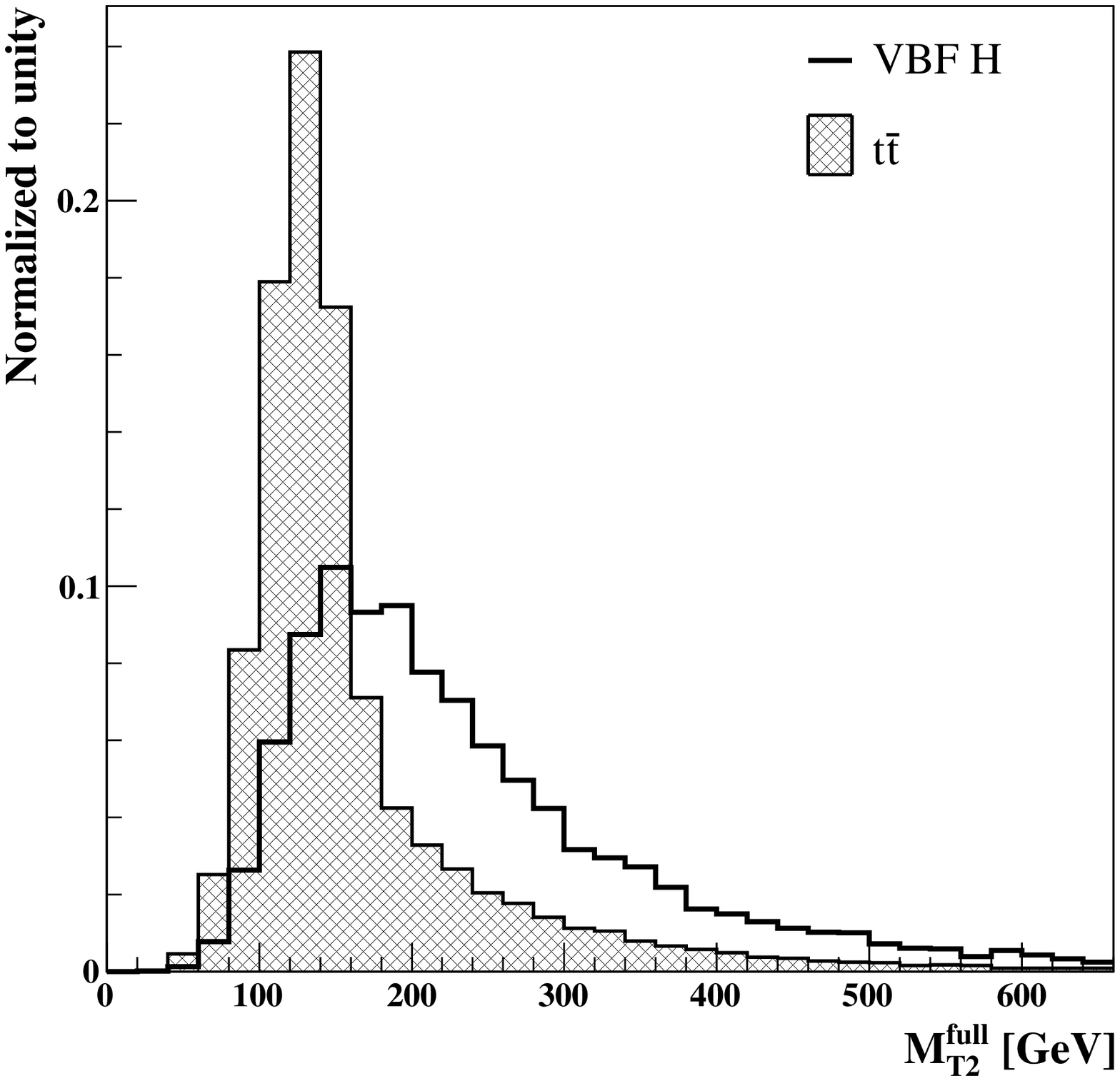, width=8.0cm, height=8.0cm}
  \end{center}
  \caption{The $M_{T2}^{\rm sub}$ (left panel) and
  $M_{T2}^{\rm full}$ (right panel) distributions of the VBF process
  ($m_H=160$ GeV)
  and the $t\bar{t}$ background, after applying
  the preselection cuts.}
  \label{fig:mt2}
\end{figure}
%

%
In the literature, the dilepton invariant mass  $M_{ll}$ and the
dijet invariant mass $M_{jj}$ for two tagging jets are also used
for the event cut.
In Fig.~\ref{fig:mllmjj}, we show  the $M_{ll}$ and $M_{jj}$
distributions for both the Higgs signal and the $t\bar{t}$
background. As in the GGF process discussed in the previous section,
one might employ $M_{T2}$  as a cut variable in the VBF process
also.
In the Higgs signal with the two tagging jets and also in the main
$t\bar{t}$ background with two $b$ jets, one can construct two sorts
of $M_{T2}$---one using the transverse masses of the $W$-pair
system and the other using the $W$-pair + two jets system:
\begin{eqnarray}
  M_{T2}^{\rm sub} &\equiv &\min_{\mathbf{k}_{T} + \mathbf{l}_{T}
  =\psl_T}
\left[\max\left\{M_T^{{\rm sub}\,(1)},\,M_T^{{\rm sub}\,(2)}
  \right\}\right]\,, \nonumber \\
  M_{T2}^{\rm full} &\equiv &\min_{\mathbf{k}_{T} + \mathbf{l}_{T}
  =\psl_T}
\left[\max\left\{M_T^{{\rm full}\,(1)},\,M_T^{{\rm full}\,(2)}
  \right\}\right]\,,
\end{eqnarray}
where $M_T^{{\rm sub}\,(i)}$ and $M_T^{{\rm full}\,(i)}$ are the
transverse masses of $W^{(i)}\rightarrow l^{(i)}\nu^{(i)}$ and
$j^{(i)}W^{(i)}\rightarrow j^{(i)}l^{(i)}\nu^{(i)}$, respectively.
The $M_{T2}^{\rm sub}$ variable is defined without any combinatorial
ambiguity and bounded from above by $\min(m_W,\,m_H/2)$ [see
Eq.~(\ref{eq:mt2max_signal})] in the signal and by $m_W$ in the
background.
On the other hand, there is a combinatorial uncertainty in
constructing $M_{T2}^{\rm full}$ since, in each selected event,
there are two possibilities in associating two jets with two charged
leptons.
Between the two possibilities,
we choose the combination giving the smaller $M_{T2}^{\rm full}$.
%
%
%
%
%
%
%
%
Then, for the $t\bar{t}$ background, $M_{T2}^{\rm full}$ is bounded
from above by $m_t$ at parton level, which suggests that
$M_{T2}^{\rm full}$ can be used as a cut variable to eliminate the
$t\bar{t}$ backgrounds by requiring $M_{T2}^{\rm full}>m_t$. For the
Higgs signal events, tagging jets tend to have a large $M_{jj}$,
which has been used to select the signals by requiring $M_{jj}>
M_{jj}^{\rm cut}$.
In Fig.~\ref{fig:mt2}, we show the $M_{T2}^{\rm sub}$ (left panel) and
$M_{T2}^{\rm full}$  (right panel) distributions for the signal and the
$t\bar{t}$ background at detector level. In the right panel of
Fig.~\ref{fig:mt2}, we observe that the $M_{T2}^{\rm
full}$ distribution for $t\bar{t}$ backgrounds has a non-negligible
tail above $m_t$, which arises from the fact that one (or both) of the
tagging dijets used for the construction of $M_{T2}^{\rm full}$ is
not the $b$ jet from the top quark decay, but a hard
jet coming from the initial state radiation or an erroneously
reconstructed jet. With this, comparing
to $M_{jj}$ in Fig.~\ref{fig:mllmjj}, we see that $M_{T2}^{\rm
full}$ can be as efficient as $M_{jj}$ in discriminating the Higgs
signal from the $t\bar{t}$ background. On the other hand,
Figs.~\ref{fig:mt2} and \ref{fig:mllmjj} show that both the Higgs
signal and the $t\bar{t}$ background have a similar shape of
$M_{T2}^{\rm sub}$ distribution, so $M_{T2}^{\rm sub}$ is not so
useful as a cut variable.
\begin{table}[t!]
  \caption{Cut flows (in fb) for $m_H = 160$ GeV.
}
  \begin{center}
    \begin{tabular}{ l|ccc|l }
      \hline\hline
      Selection cuts & VBF $H\rightarrow WW$ & $t\bar{t}$ & EW $WW+$ jets &$S/B$ \\
      \hline
      Preselection & 33.8 & 10910.0 & 15.9 & 0.0031\\
      $b$-jet veto & 32.8 & 5193.1 & 15.5 & 0.0063\\
      Forward jet tagging & 22.5 & \,\,\,\,542.1 & \,\,6.3 & 0.041\\
      $~~$Leptons between jets
      & 20.9 & \,\,\,\,347.3 & \,\,5.6 & 0.059\\
      $~~$Central jet veto
      & 16.8 & \,\,\,\,166.9 & \,\,3.5 & 0.099\\
      $m_T^{ll\nu}>30$ GeV, $Z\rightarrow\tau\tau$ rej.
      & 15.1 & \,\,\,\,138.5 & \,\,2.7 & 0.11\\
      $M_{ll}<85$ GeV & 14.9 & \,\,\,\,\,\,62.0 & \,\,0.8 & 0.24 \\
      $M_{jj}>550$ GeV & \,\,9.3 & \,\,\,\,\,\,\,\,8.0 & \,\,0.7 & 1.07 \\
      \hline
      $\Delta\Phi_{ll}<1.6$, $\Delta\eta_{ll}<1.5$
      & \,\,8.4 & \,\,\,\,\,\,\,\,5.5 & \,\,0.5 & 1.39 \\
      \hline
      $M_{T2}^{\rm full}>160$ GeV
      & \,\,8.9 & \,\,\,\,\,\,\,\,7.1 & \,\,0.7 & 1.14 \\
      $M_{T2}^{\rm sub}>60$ GeV
      & \,\,3.6 & \,\,\,\,\,\,\,\,1.7 & \,\,0.3 & 1.80 \\
      $\Delta\Phi_{ll}<1.6$, $\Delta\eta_{ll}<1.5$
      & \,\,3.6 & \,\,\,\,\,\,\,\,1.6 & \,\,0.2 & 2.00 \\
      \hline\hline
    \end{tabular}
  \end{center}
\label{tab:cutflow160}
\end{table}
\begin{table}[t!]
  \caption{Cut flows (in fb) for $m_H = 140$ GeV.
}
  \begin{center}
    \begin{tabular}{ l|ccc|l }
      \hline\hline
      Selection cuts & VBF $H\rightarrow WW$ & $t\bar{t}$ & EW $WW+$ jets &$S/B$ \\
      \hline
      Preselection & 17.6 & 10910.0 & 15.9 & 0.0016 \\
      $b$-jet veto & 17.3 & \,\,5193.1 & 15.5 & 0.0033 \\
      Forward jet tagging & 12.2 & \,\,\,\,542.1 & \,\,6.3 & 0.022 \\
      $~~$Leptons between jets & 11.6 & \,\,\,\,347.3 & \,\,5.6 & 0.033 \\
      $~~$Central jet veto & \,\,9.8 & \,\,\,\,166.9 & \,\,3.5 & 0.058 \\
      $m_T^{ll\nu}>30$ GeV, $Z\rightarrow\tau\tau$ rej.
      & \,\,8.5 & \,\,\,\,138.5 & \,\,2.7 & 0.060\\
      $M_{ll}<85$ GeV & \,\,8.5 & \,\,\,\,\,\,62.0 & \,\,0.8 & 0.14 \\
      $M_{jj}>550$ GeV & \,\,5.7 & \,\,\,\,\,\,\,\,8.0 & \,\,0.7 & 0.66 \\
      \hline
      $\Delta\Phi_{ll}<1.7$, $\Delta\eta_{ll}<1.6$
      & \,\,5.4 & \,\,\,\,\,\,\,\,5.5 & \,\,0.5 & 0.89 \\
      \hline
      $M_{T2}^{\rm full}>140$ GeV
      & \,\,5.5 & \,\,\,\,\,\,\,\,7.5 & \,\,0.7 & 0.67 \\
      $M_{T2}^{\rm sub}>45$ GeV
      & \,\,2.1 & \,\,\,\,\,\,\,\,3.7 & \,\,0.5 & 0.50 \\
      $M_{T2}^{\rm sub}<72$ GeV
      & \,\,2.1 & \,\,\,\,\,\,\,\,3.1 & \,\,0.3 & 0.62 \\
      $\Delta\Phi_{ll}<1.7$, $\Delta\eta_{ll}<1.6$
      & \,\,2.1 & \,\,\,\,\,\,\,\,2.6 & \,\,0.3 & 0.72 \\
      \hline\hline
    \end{tabular}
  \end{center}
\label{tab:cutflow140}
\end{table}
With the above observations,  in our event selection we have applied
$M_{ll}$ and $M_{jj}$ cuts after the basic selection cuts:
\begin{equation}
M_{ll}\,<\, M_{ll}^{\rm cut}, \quad M_{jj}\,>\, M_{jj}^{\rm
cut}.\end{equation}
Then, more refined $M_{T2}$ cuts have been introduced
to reduce the number of background events further. In
Tables~\ref{tab:cutflow160} and \ref{tab:cutflow140}, we show how
the cross sections of the signal and backgrounds change under each
selection cut for $m_H=160$ GeV and 140 GeV, respectively.
We note that one may also use the upper cut on the $M_{T2}^{\rm
sub}$ when $m_H\leq 2m_W$; see the third set of cuts in
Table~\ref{tab:cutflow140}. This is because the signal distribution
is bounded above by $m_H/2<m_W$ in this case. This upper cut cannot
be applied when $m_H\geq 2m_W$, as the $M_{T2}^{\rm sub}$
distributions of both the signal and the background have an equal
maximum, $m_W$.

\begin{figure}[t!]
  \begin{center}
{\epsfig{figure=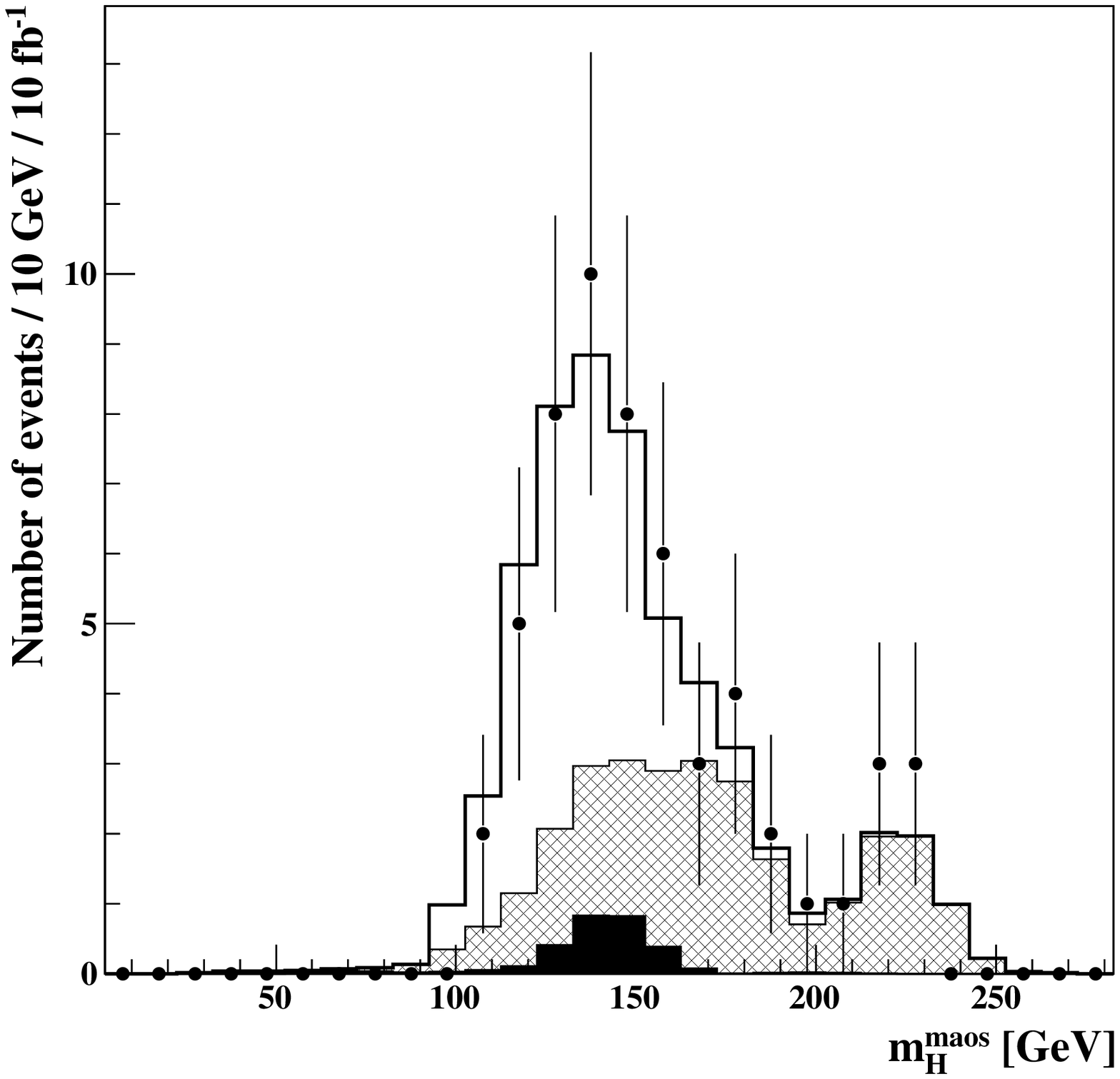,height=8.0cm,width=8.0cm}}
{\epsfig{figure=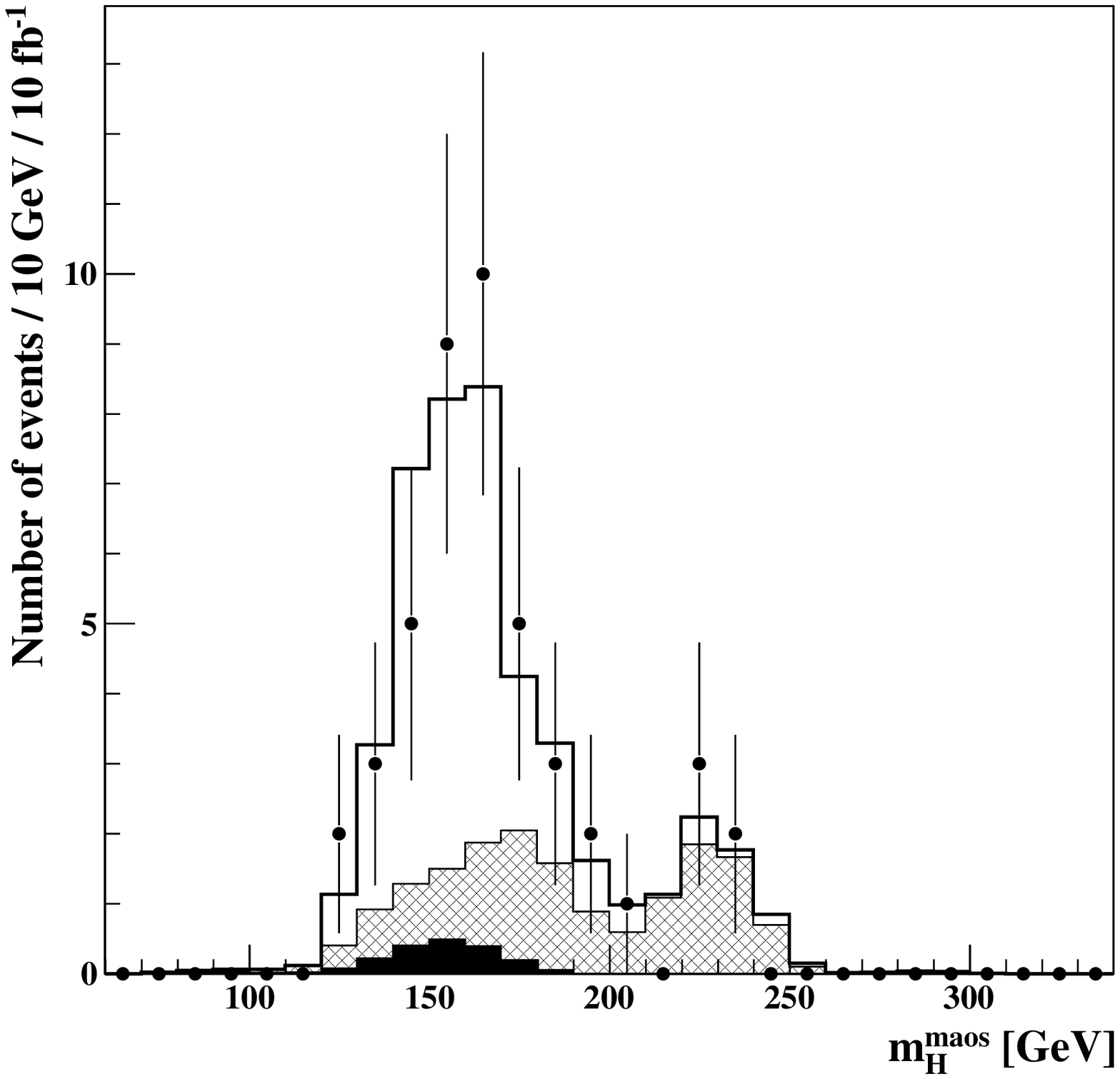,height=8.0cm,width=8.0cm}}
  \end{center}
\caption{The $m_H^{\rm maos}$ distributions for the
pseudoexperiment data (dots) and the template (solid line) for
$m_H=$ 140 (left) and 160 GeV (right) 
at 10 fb$^{-1}$.
The lightly and thickly
shaded regions represent the $t\bar{t}$ and $WW+2j$ backgrounds,
respectively.} \label{fig:mhmaos_VBF}
\end{figure}

In Fig.~\ref{fig:mhmaos_VBF}, we show the ${m}_H^{\rm maos}$
distributions for the two nominal values of the Higgs boson mass:
$m_H=140$ and 160 GeV. Here $\left(m_H^{\rm maos}\right)^2
=\left(p_1+k_1^{\rm maos}+p_2+k_2^{\rm maos}\right)^2$ for
$W^{(1)}W^{(2)}\rightarrow
l^{(1)}(p_1)\nu^{(1)}(k_1)l^{(2)}(p_2)\nu^{(2)}(k_2)$, where the
transverse components of $k_i^{\rm maos}$ ($i=1,2$) are those used to
determine $M_{T2}^{\rm sub}$, and the longitudinal and energy
components are fixed by the constraints $\left(p_1+k_1^{\rm
maos}\right)^2= \left(p_2+k_2^{\rm maos}\right)^2=(M_{T2}^{\rm
sub})^2$ and $\left(k_1^{\rm maos}\right)^2=\left(k_2^{\rm
maos}\right)^2=0$.
We observe that the
$m_H^{\rm maos}$ distribution has a clear peak at the true (nominal)
Higgs mass.\footnote{
The background bump appearing around $m_H^{\rm maos} \gsim 200$ GeV is 
an accidental consequence of our analysis for the VBF case. We observe 
that it becomes less significant like as in the GGF case for a 
different choice of the bin size.
}
With the ${m}_H^{\rm maos}$ distribution constructed as above, we
performed a template fitting to determine the Higgs boson mass as in
Sec.~\ref{sec:GF}, the GGF case. The estimated Higgs
masses, together with the 1-$\sigma$ errors obtained by fitting the log
likelihood distributions for various Higgs masses, are listed in
Table~\ref{tab:mh_fitted}. The 1-$\sigma$ deviated value is defined
as the one that increases $-\ln\mathcal{L}$ by
$1/2$~\cite{Amsler:2008zzb}. 
We again see that the
reconstructed mass is quite close to the input Higgs boson mass. 

%
%
\begin{table}[t!]
\caption{
The Higgs boson masses and the 1-$\sigma$
statistical errors obtained by the likelihood fit to the $m_H^{\rm
maos}$ distributions in the VBF processes at 10 fb$^{-1}$. 
}
  \begin{center}
    \begin{tabular}{ c | c c c c c c c }
    \hline \hline
    $m_H$ (GeV) & 130 & 140 & 150 & 160 & 170 & 180 & 190 \\
    \hline
    Fitted value (GeV) & 132.5 & 140.7 & 150.8 & 161.0 & 170.6 & 183.4 & 188.6 \\
    1-$\sigma$ error (GeV) & ~~~7.5 & ~~~7.0 & ~~~4.8 & ~~~3.7 & ~~~3.8 & ~~~6.2 & ~~~6.8 \\
    \hline \hline
    \end{tabular}
  \end{center}
\label{tab:mh_fitted}
\end{table}

\begin{figure}[t!]
  \begin{center}
{\epsfig{figure=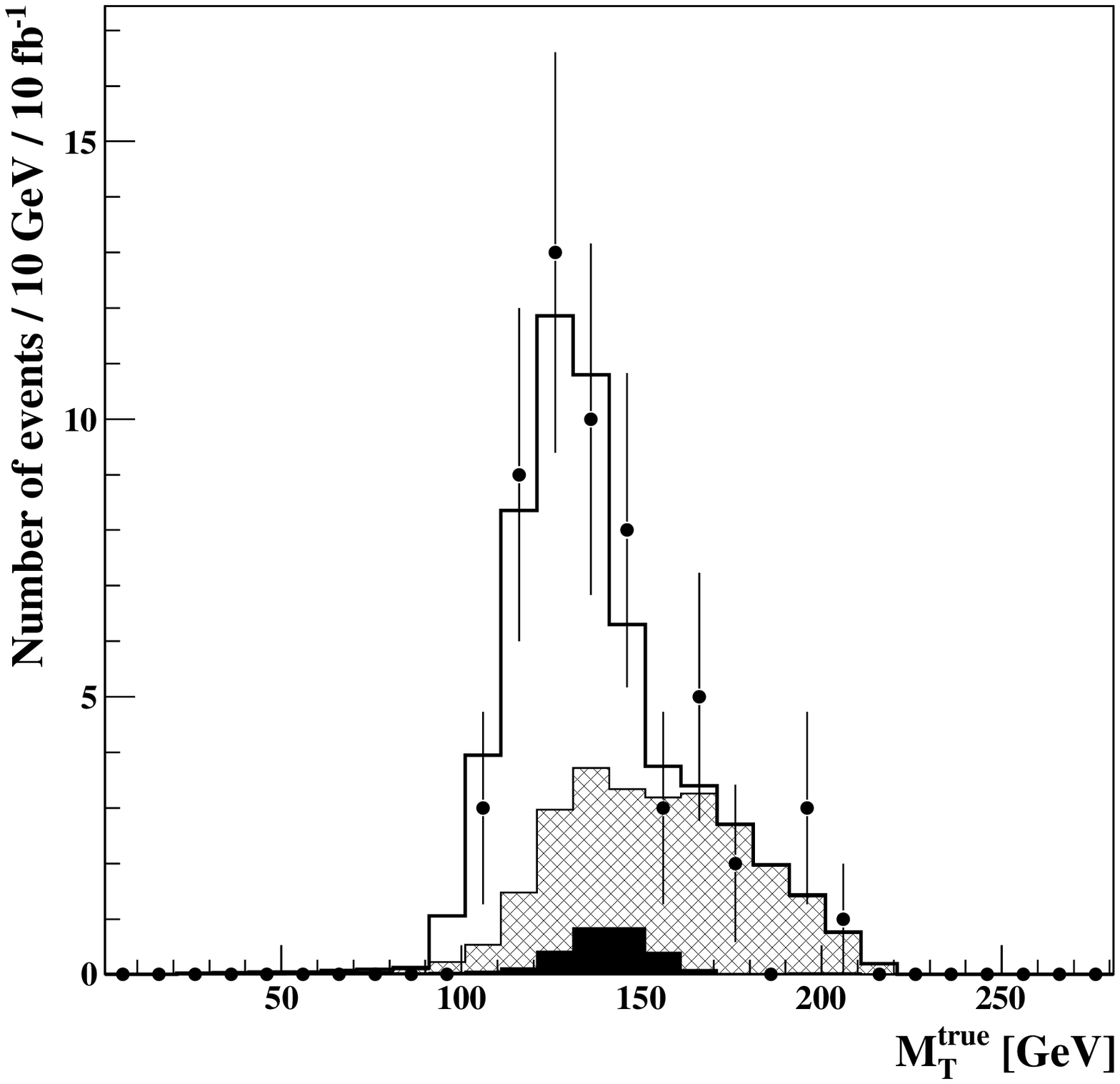,height=8.0cm,width=8.0cm}}
{\epsfig{figure=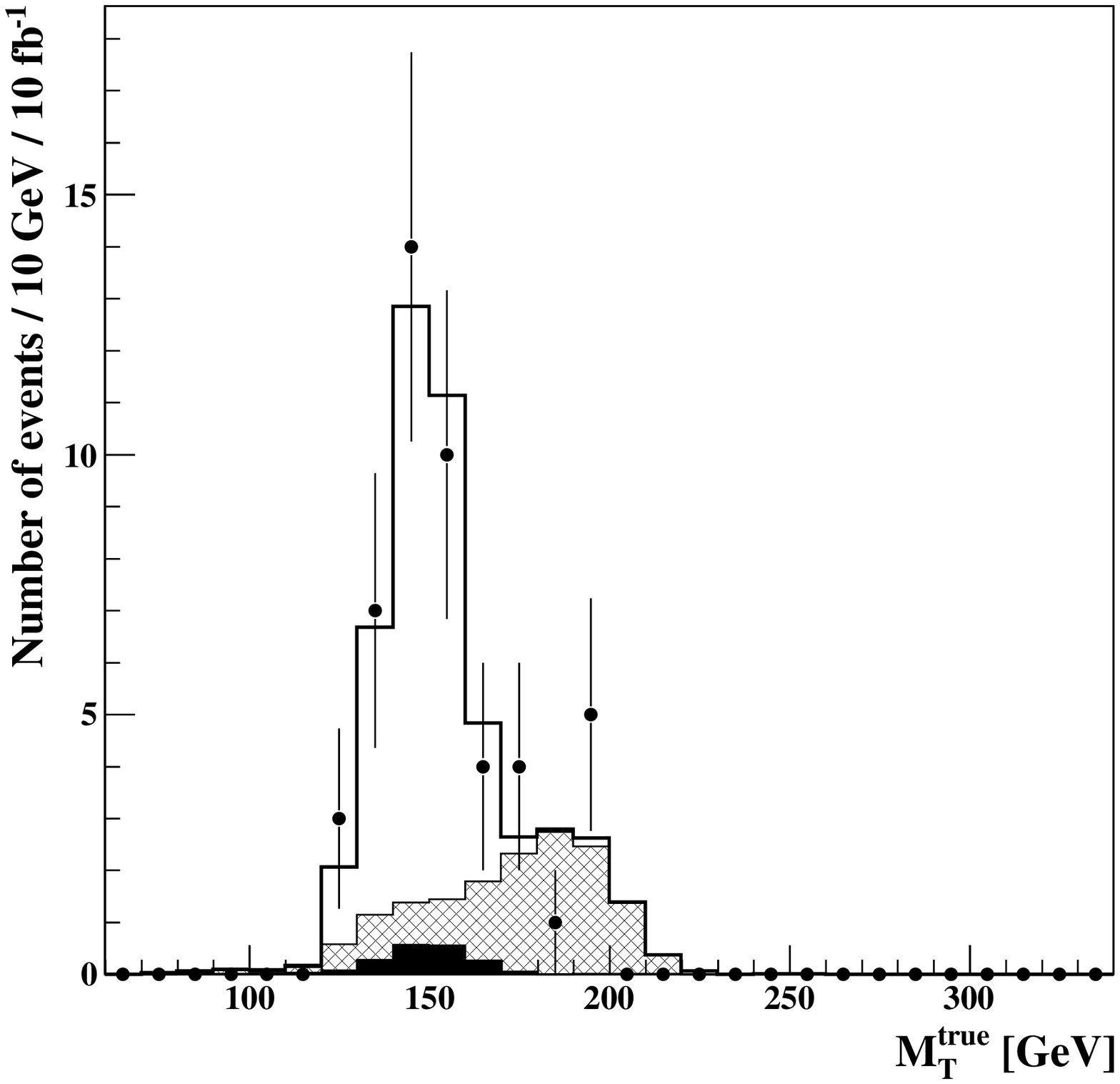,height=8.0cm,width=8.0cm}}
  \end{center}
\caption{The $M_T^{\rm true}$ distributions for the
pseudoexperiment data (dots) and the template (solid line) for
$m_H=$ 140 (left) and 160 GeV (right) 
at 10 fb$^{-1}$ luminosity.
The lightly and thickly shaded regions
represent the $t\bar{t}$ and $WW+2j$ backgrounds, respectively.}
\label{fig:mTtrue_VBF}
\end{figure}
\begin{figure}[t!]
\begin{center}
\epsfig{figure=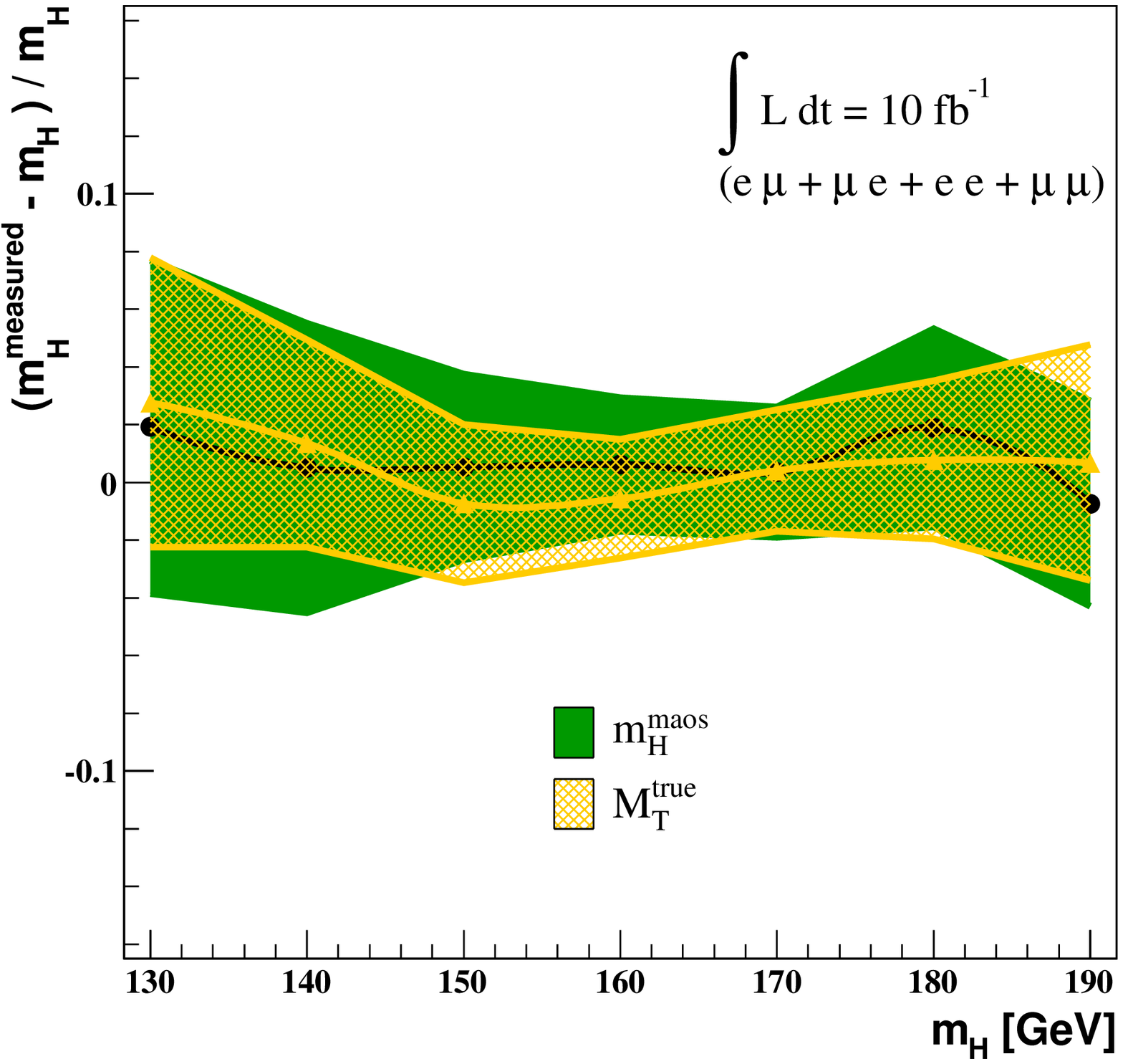,height=8.0cm,width=8.0cm}
\epsfig{figure=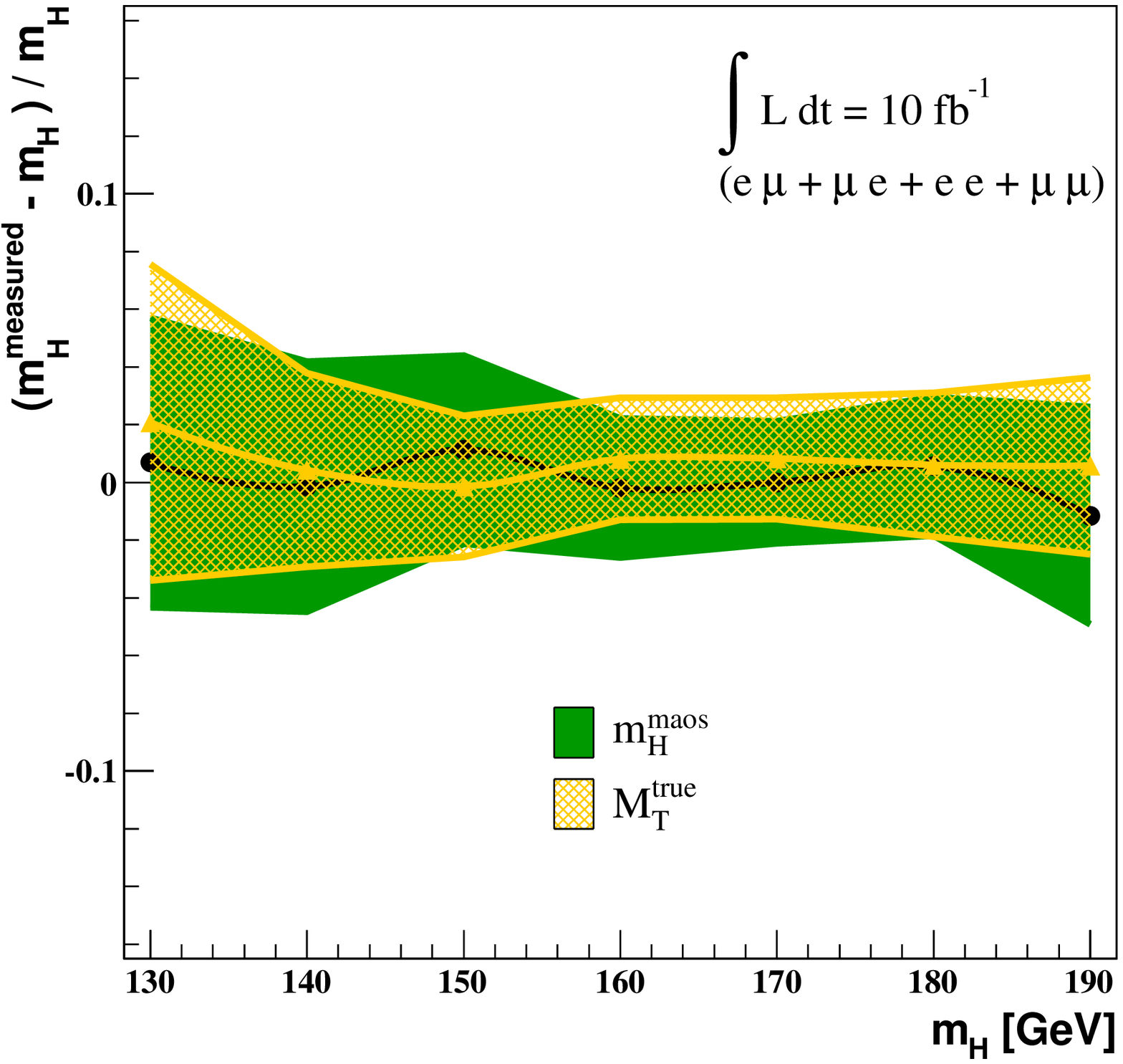,height=8.0cm,width=8.0cm}
\end{center}
\caption{
(Left panel) The band showing the 1-$\sigma$ deviation error for the
Higgs boson mass determined by the $m_H^{\rm maos}$
and $M_T^{\rm true}$ distributions in the VBF process
$qq\to qqH$ with
$H \to WW^{(*)} \to l\nu \, l^\prime\nu^{\,\prime}$.
The dots and lines denote the Higgs boson mass obtained by the
likelihood fit to the pseudoexperiment distributions.
(Right panel) The same as in the left frame
but without the $M_{T2}$ cuts.
}
\label{fig:VBFcomparison}
\end{figure}
%

For a comparison of the efficiency of $m_H^{\rm
maos}$ to that of $M_T^{\rm true}$, we also perform a likelihood fit
of the $M_T^{\rm true}$ distribution with the same event sets,
including the backgrounds.
In Fig.~\ref{fig:mTtrue_VBF}, we show the $M_T^{\rm true}$ template
distributions (solid lines) when the trial Higgs mass is 140
GeV (left panel) and 160 GeV (right panel). The dots with error bars represent
the $M_T^{\rm true}$ distributions for the pseudoexperiment data
with the same nominal Higgs masses.
We show our results on the Higgs boson masses and the 1-$\sigma$ errors
in the left frame of Fig.~\ref{fig:VBFcomparison};
these results are obtained using the likelihood fit to
the $m_H^{\rm maos}$ and $M_T^{\rm true}$ distributions
in the VBF process
$qq\to qqH$ with
$H \to WW^{(*)} \to l\nu \, l^\prime\nu^{\,\prime}$.
We observe that the efficiencies of the two methods are comparable
to each other, and both of them provide a good accuracy for the 
determination of the Higgs mass.
Furthermore, to see the role of the $M_{T2}$ cuts in the measurement,
we repeat the likelihood fitting of the distributions after applying
the same event selection cuts except the $M_{T2}$ cuts; see the
right frame of Fig.~\ref{fig:VBFcomparison}.  
We find that the final result does not change much since the purity of the
event samples is already high enough before imposing the $M_{T2}$
cuts and also the number of signals is relatively smaller than that
of the GGF process.
Although $M_T^{\rm true}$ and $m_H^{\rm maos}$ have a similar
efficiency for the VBF process, we again note that in a situation
where there are unknown backgrounds beyond the SM which might smear
the endpoint of the $M_T^{\rm true}$ distribution, the $m_H^{\rm
maos}$ could be a better choice  since the peak is less vulnerable
to unknown backgrounds than the endpoint.

%

\section{Conclusions}
\label{sec:concl}
%
We performed  a comparative study of the Higgs boson
mass measurements based on two kinematic observables, $M_T^{\rm
true}$, and $m_H^{\rm maos}$ in dileptonic decays of a $W$ boson pair.
For this, we first discussed some features of $M_T^{\rm true}$ and
$M_{T2}$, and also of the 
MAOS
reconstruction of the neutrino momenta in dileptonic $W$ boson
decays and the associated invariant mass observable $m_H^{\rm
maos}$. It is found that $M_T^{\rm true}$ and $m_H^{\rm maos}$ can
determine the Higgs boson mass in the range 130 GeV $\leq\,
m_H\,\leq$ 190 GeV with a similar accuracy for both of the two main
production mechanisms of the SM Higgs boson at the LHC, i.e.
GGF and VBF,
up to systematic errors which are not considered in our analysis.
Still, it should be noted that the $m_H^{\rm maos}$ distribution has
a peak at $m_H$, while the  $M_T^{\rm true}$ distribution has an
endpoint, and thus $m_H^{\rm maos}$ can be a better observable when
there are some additional backgrounds due to new physics beyond the
SM since it is likely that the peak is less vulnerable to unknown
backgrounds than the endpoint.
One might consider an approach using both $m_H^{\rm
maos}$ and $M_T^{\rm true}$ together to extract maximal information
on the Higgs boson mass from experimental data. However, our results
in Sec.~\ref{sec:theo} imply that these two variables have a
rather strong correlation, and therefore it is not likely that such an
approach significantly improves the accuracy of the Higgs mass
measurement.

Our study suggests that $M_{T2}$ can be a useful cut variable in the GGF
process, as the signal distribution becomes narrower while the
background distribution becomes flatter  under an $M_{T2}$ cut.
Although systematic errors are not considered in our analysis,
such behavior of the signal and background distributions might be
useful, particularly for reducing various systematic uncertainties
in the real analysis of experimental data. In our analysis of the VBF
process, $M_{T2}$ is not a particularly useful cut variable compared
to others such as the dijet and dilepton invariant masses.

\vspace{-0.2cm}
\subsection*{Acknowledgements}
\vspace{-0.3cm} We thank W.~S.~Cho for useful
discussions. This work was supported by the KRF Grants funded by
the Korean Government (No. KRF-2008-314-C00064 and No. KRF-2007-341-C00010)
and the KOSEF Grant funded by the Korean Government (No.
2009-0080844). \noindent


%

\end{document}